%% file: main_sigmetrics.tex
\documentclass[acmsmall,screen]{acmart}
\AtBeginDocument{%
  }

\setcopyright{cc}
\setcctype{by}
\acmJournal{POMACS}
\acmYear{2025} \acmVolume{9} \acmNumber{3} \acmArticle{61} \acmMonth{12} \acmPrice{}\acmDOI{10.1145/3771576}
\usepackage{tikz}
\usepackage{amsmath}
\usepackage{mathtools}
\usepackage{amsthm}
\usepackage{amsfonts}
\usepackage{bm}
\usepackage{gensymb}
\usepackage{ulem}
\usepackage{xspace}
\usepackage{subfig} 
\usepackage{enumitem}
\usepackage{verbatim}
\usepackage{subcaption}
\usepackage{graphicx}
\usepackage{caption}
\usepackage{multirow}
\usepackage{hyperref}
\usepackage{subfloat}
\usepackage{bookmark}

\usepackage{blindtext}

\newcommand{\sys}{{\scshape SageServe}\xspace} 

\newcommand{\csp}{{Microsoft O365}\xspace}

\usepackage[most]{tcolorbox}

\newtcolorbox{mytextbox}[1][]{%
  sharp corners,
  enhanced,
  colback=white,
  attach title to upper,
  #1
}

\definecolor{revcolor}{rgb}{0.8,0.2,0.2}
\definecolor{yscolor}{rgb}{0.7,0.3,0.7}

\newcommand{\para}[1]{{\textsf{\textit{#1.~}}}}

\usepackage{pifont}
\usepackage[english]{babel}
\addto\extrasenglish{
  
}
\addto\extrasenglish{
  
}
\addto\extrasenglish{
  
}

\newcommand{\colortakeaway}[1]{%
  \noindent\colorbox{blue!15}{\parbox{\dimexpr\columnwidth-0\fboxsep}{#1}}}

\begin{document}

\title{\sys: Optimizing LLM Serving on Cloud Data Centers with Forecast Aware Auto-Scaling}

\author{Shashwat Jaiswal}
\authornote{Equal contribution}
\email{sj74@illinois.edu}
\affiliation{%
  \institution{University of Illinois Urbana-Champaign}
  \country{USA}
}
\authornote{Work done as an intern at Microsoft}

\author{Kunal Jain}
\authornotemark[1]
\email{kjain324@gatech.edu}
\affiliation{%
  \institution{Georgia Institute of Technology}
  \country{USA}
}
\authornote{Work done as a Research Fellow at Microsoft}

\author{Yogesh Simmhan}
\email{simmhan@iisc.ac.in}
\affiliation{%
    \institution{Indian Institute of Science}
    \city{Bangalore}
  \country{India}
}

\sloppy \author{Anjaly Parayil, Ankur Mallick, Rujia Wang, Renee St. Amant, \mbox{Chetan Bansal}, Victor Ruhle, Anoop Kulkarni, Steve Kofsky, \mbox{Saravan Rajmohan}}
\email{aparayil@microsoft.com}
\affiliation{%
    \institution{Microsoft}
  \country{India, UK and USA}
}

\renewcommand{\shortauthors}{Shashwat Jaiswal et al.}

\begin{abstract}
Global cloud service providers handle inference workloads for Large Language Models (LLMs) that span latency-sensitive (e.g., chatbots) and insensitive (e.g., report writing) tasks, resulting in diverse and often conflicting Service Level Agreement (SLA) requirements. Managing such mixed workloads is challenging due to the complexity of the inference serving stack, which encompasses multiple models, GPU hardware, and global data centers. Existing solutions often silo such fast and slow tasks onto separate GPU resource pools with different SLAs, but this leads to significant under-utilization of expensive accelerators due to load mismatch.
In this article, we characterize the LLM serving workloads at Microsoft Office 365, one of the largest users of LLMs within Microsoft Azure cloud with over {10 million requests per day}, and highlight key observations across workloads in different data center regions and across time. This is one of the first such public studies of Internet-scale LLM workloads.
We use these insights to propose \sys, a comprehensive LLM serving framework that dynamically adapts to workload demands using multi-timescale control knobs. It combines short-term request routing to data centers with long-term scaling of GPU VMs and model placement with higher lead times, and co-optimizes the routing and resource allocation problem using a traffic forecast model and an Integer Linear Programming (ILP) solution.
We evaluate \sys through real runs and realistic simulations on {10 million production requests}
across three regions and four open-source models. We achieve up to 25\% savings in GPU-hours compared to the current baseline deployment and reduce GPU-hour wastage due to inefficient auto-scaling
by 80\%, 
resulting in a potential monthly cost savings of up to \$2.5 million, 
while maintaining tail latency and meeting SLAs.
The workload traces, our simulator harness and the \sys scheduler are available at \url{https://github.com/shashwatj07/SageServe}.
\end{abstract}




\received{July 2025}
\received[revised]{September 2025}
\received[accepted]{October 2025}

\maketitle

\input{SC_25/introduction}

\input{SC_25/SystemandApplication}
\input{SC_25/Optimization}
\input{SC_25/architecture}
\input{SC_25/evaluation}
\input{SC_25/RelatedWork}
\input{SC_25/Conclusions}

\section{Acknowledgements}
We would like to thank all the reviewers for their constructive feedback and our shepherd, Prof. Jian Li for helping us incorporate all the reviews and improving the final version of our work.

\clearpage
\bibliographystyle{ACM-Reference-Format}
\bibliography{main}

\input{sigmetrics/Appendix}


\end{document}

%% file: SC_25/introduction.tex
\section{Introduction}\label{sec:intro}
\para{Motivation}
Recent years have seen rapid adoption of \textit{Large Language Models (LLMs)} in enterprise products and services to power both proactive and user-initiated intelligent features~\cite{Bommasani2021FoundationModels}. As their capabilities expand, LLM usage is growing exponentially across enterprise, consumer, and scientific applications~\cite{zhao2025llm,cui2025curieevaluatingllmsmultitask,lu2024aiscientistfullyautomated}.The growth of Agentic AI and workflows is only accelerating this, with LLM agents enabled with tool execution autonomously completing complex tasks~\cite{murugesan2025rise,zhang2025aflow,he2025llm}.

Cloud-hosted, GPU-accelerated Virtual Machines (VMs) are central to scaling LLM inference, prompting major investments from Cloud Service Providers (CSPs) for both internal use and public offerings. 
AWS UltraClusters offer 100,000 of Trainium2 accelerators used by Anthropic and end-users~\cite{hpc_wire} while Microsoft Azure Cloud's Eagle system with H100 GPUs features at \#5 in the Top500 supercomputing list~\cite{top500} and Google has introduced the top-end NVIDIA HGX B200 GPUs into its data centers~\cite{gcp_update}.
However, maximizing return on these expensive resources is critical. Misalignment between GPU provisioning and traffic distribution across regions can lead to Service Level Agreement (SLA) violations or resource wastage at either extremes. The need to meet SLAs and provide a smooth user experience often causes CSPs to over-provision GPU capacity, raising infrastructure costs, increasing prices for users, and diverting resources from R\&D~\cite{meta-llm-serving,miao2023towards}.
This happens at multiple levels: from routing user requests across data centers, to scaling the VM instances and models, and routing within a region to meet the demand~\cite{stojkovic2025tapas,fu2024serverlessllm,jain2025performance}, to optimizing the execution for a single model instance across GPUs~\cite{gao2024cost,wu2024loongserve,patel2024splitwise}. We focus on the former problems.

\para{Challenges}
Unlike VM and container auto-scaling for traditional cloud workloads~\cite{wang2021faasnet, hadary2020protean,rzadca2020autopilot}, scaling GPU VMs for LLM workloads presents unique challenges. Commercial platforms like Google Gemini~\cite{Gemini}, Microsoft Copilot~\cite{Copilot}, and OpenAI ChatGPT~\cite{ChatGPT} serve a mix of models and inference \textit{workload tiers} that can broadly be categorized as: (a) \textit{Interactive workload (IW)} that are latency-sensitive and require real-time responses, within seconds, e.g., chatbots, LLM powered search, content moderation; and (b) \textit{Non-interactive Workload (NIW)} requests, which are less time-critical and focus on serving resource-intensive or batch processes, e.g., report writing, data annotation, and simulations within 10s of minutes or hours. These workloads and their priority vary by time, region, and user type (enterprise/consumer), making it difficult to design a unified auto-scaling policy that efficiently handles diverse models and SLA requirements.

\begin{figure}[t]
\vspace{-0.1in}
    \centering
\includegraphics[width=0.8\columnwidth]{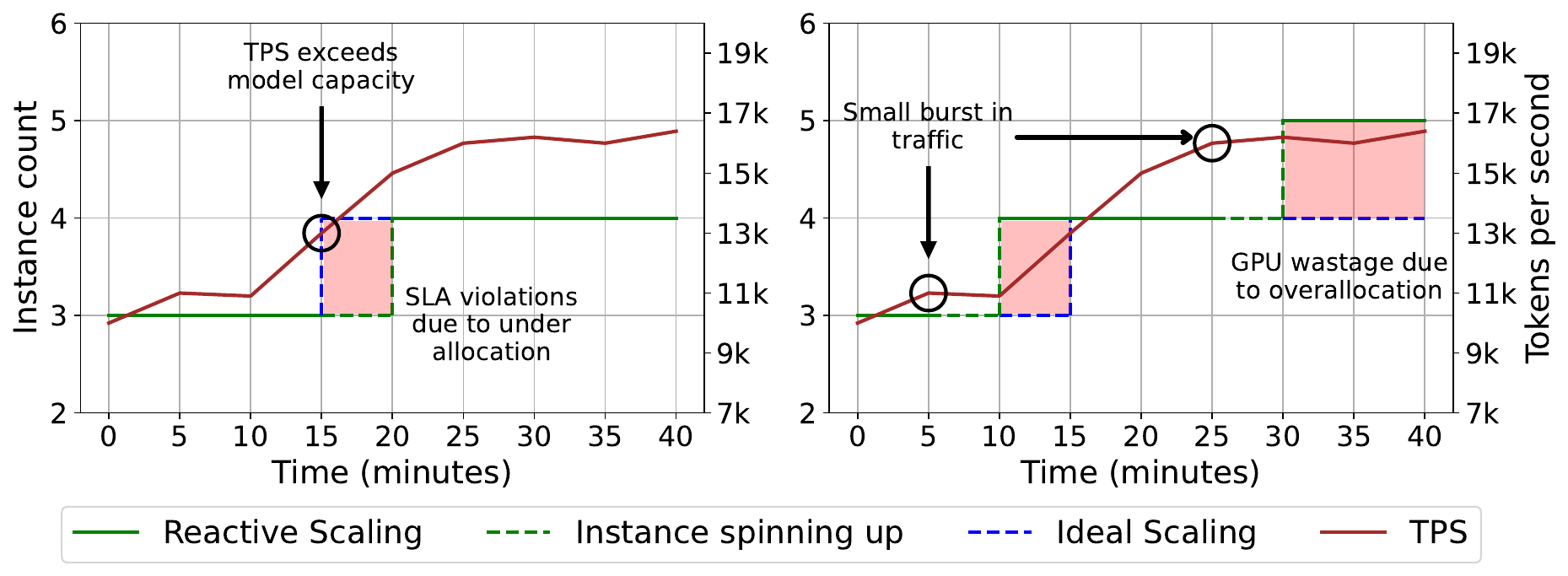}
\vspace{-0.1in}
    \caption{VM instance scaling (left Y axis) based on incoming TPS (right Y axis) using ideal (pink) and reactive (green) strategies. Shaded region shows the difference in their instance counts.
    }
    \label{fig:ideal_scaling}
    \vspace{-0.1in}
\end{figure}

Dynamically scaling LLM model instances just-in-time can be \textit{ineffective}, due to traffic variations, and \textit{slow}, blocking GPUs for many seconds or minutes during cold starts, when loading large LLMs~\cite{aws_fast_loader}, e.g., Meta's Llama2-70B model~\cite{Llama} is $\approx140$GB in FP16, and OpenAI's GPT models are estimated to be even larger. Reactive scaling that instantiates new VMs and model instances based on real-time metrics, such as incoming \textit{Tokens Processed per Second (TPS)}, can cause over- or under-provisioning if they fail to account for TPS variance and LLM loading delays. E.g., \autoref{fig:ideal_scaling} illustrates a scenario where the model instance has a capacity to serve $4000$ TPS. In the first plot, the reactive strategy decides to scale up the instance count at $T=15$~mins, which causes the instance become available only at $T=20$~mins dur to cold start, resulting in SLA violations for $5$~mins due to \textit{under-allocation}. If we had use a conservative instance capacity of $3500$ TPS, the reactive approach is susceptible to \textit{over-allocation}, as seen in the second plot, where we unnecessarily scale up at $T=25$~mins due to a small increase in traffic even though the input TPS later stabilizes. 

There is a pressing need for a flexible, lightweight auto-scaling policy that adapts to dynamic workloads, minimizes model loading delays, and reduces costs while meeting diverse SLAs. Besides a knowledge of the LLM serving infrastructure and pipeline design, this also requires access to workload traces and an analysis of their characteristics.

\para{Gaps}
Meta~\cite{meta-llm-serving} discuss the presence of daily peaks, off-peaks, and unpredictable spikes in LLM inference workloads, and validates the presence of both IW and NIW.
They emphasize that achieving cost-effective solutions at scale requires extensive benchmarking and production-level insights.
However, 
a lack of public production traces causes literature to often rely on synthetic or regionally scoped datasets that lack key attributes~\cite{patel2024splitwise, agrawal2024vidur, wang2024burstgpt}.
In this work, we extensively characterize cloud-scale LLM workloads with different SLAs, and will also place the traces in the public domain.
While there are routing and scaling strategies proposed for general workloads and LLMs, they make simplifying assumptions.  Jain et al.~\cite{jain2025performance} assume identical LLM types and workloads of equal priority, focusing on load balancing across multiple instances within a region. TAPAS~\cite{stojkovic2025tapas} is also orthogonal, focusing on the power and thermal characteristics of LLM workloads in a region. 
Others~\cite{chiron, fu2024serverlessllm} leverage additional storage and memory capacity for faster model loading and live request migration. Chiron~\cite{chiron} also explores mixing interactive and non-interactive requests and proposes various instance scaling solutions.
However, these focus on regional-level deployments, overlooking inter-region imbalances and the load disparities across LLM types within a region. We address these macro challenges across regions as well, to holistically to  utilize available capacity.
Model optimization strategies~\cite{patel2024splitwise,gao2024cost,wu2024loongserve} also complement our work at the model instance level.

\para{Approach}
We first perform a \textit{principled characterization} of real-world LLM inferencing workloads at Microsoft's Office 365 group (O365), which is one of the largest users of LLMs within Microsoft Azure through its Copilot capabilities, serving over {$\approx 10 M$ LLM requests per day} across the major US data centers. We use diverse metrics to highlight the spatial and temporal features, and predictable request patterns in these workloads for key LLM models. We use both recent (Jul, 2025) and past (Nov, 2024) traces from $3$ US data centers, charting the evolution of these workloads. We also identify differing SLAs based on latency and priority for the top inferencing applications, and segregate them into Interactive Fast (IW-F) and Normal (IW-N), and Non-Interactive (NIW) workloads.

We then describe the \textit{current inference serving design}, consisting of request routing to regions, routing to a model instance endpoint within that region, and then their local execution on a model deployment (\autoref{fig:simulator}). We highlight the limitations of the current \textit{siloed approach} (\autoref{fig:arch_silo}), 
where separate GPU pools are maintained for IW and NIW requests, which leads to significant under-utilization of the IW resources during off-peak hours. 
This motivates the design of our proposed scheduler, \sys, that introduces a \textit{reactive heuristic} over a unified pool of GPU resources that are shared by all workload types (\autoref{fig:arch_react}). It intelligently queues and
releases NIW requests to under-loaded GPUs to save GPU-hours while meeting SLAs. 
To address mismatches between the \textit{inferencing request load} and the available capacity of different LLM \textit{model instances}
(e.g., GPT-4.1, Llama4-Scout), \sys then applies a \textit{predictive heuristic} (\autoref{fig:arch_predict}), formulating a constrained optimization problem for scaling LLM instances based on ARIMA-based times-series forecasts of inference requests~\cite{shumway2017arima}. 
Subsequently, a \textit{reactive scaling heuristic} based on GPU memory utilization -- a proxy for load -- is used to address minor traffic fluctuations.
These improve GPU utilization, ensure SLA compliance to maintain tail latency, and allow surplus capacity from O365 to be leveraged for additional LLM services, e.g., Azure OpenAI Service.
\sys is validated using a realistic simulation harness that we have developed by extending SplitWise~\cite{patel2024splitwise}, using 4 popular open-source LLMs, for week-long traces from multiple US data centers across two time periods, and also compared against Chiron~\cite{chiron}, a State of the Art (SOTA) baseline.

Our characterization study is unique in examining large-scale LLM inferencing workloads in an operational system, highlighting opportunities for optimizing such serving pipelines. \sys is designed for real-world deployment, intelligently leveraging predictive scaling and scheduling strategies to meet the SLAs, and delivering practical benefits.

\para{Contributions}
We make the following specific contributions in this paper:
\begin{enumerate}[leftmargin=*,noitemsep,topsep=0pt,parsep=0pt,partopsep=0pt]
\item We describe the current LLM serving platform at \csp, which routes requests and manages GPU VMs in siloed resource pools across regions to handle diverse workloads  (\autoref{sec:sys-app-model}). We then offer a principled characterization of these workload tiers using diverse metrics for the top IW and NIW applications in US regions (\autoref{sec:wlstudy}).

\item  
We highlight the resource inefficiencies of siloed resources for workload tiers, which motivates the need for a systematic approach to GPU VM and LLM instance scaling with continuous optimization (\autoref{sec:empstudy}). 
We propose \sys, a unified framework to serve diverse LLM inference workloads across cloud regions, aiming to meet SLAs while maximizing GPU resource efficiency. We formally specify this as an 
optimization problem (\autoref{sec:optimization}), and leverage traffic forecasting models and consider practical factors such as instance capacity (in TPS), IW demand, NIW headroom, provisioning overheads, and surplus donations to public services in our design (\autoref{sec:arch}).
\item We implement a prototype of \sys and evaluate it using a realistic simulation harness based on SplitWise~\cite{patel2024splitwise}. We compare \sys with a SOTA auto-scaler, \textit{Chiron}~\cite{chiron}, for two week-long real-world traces {(10M requests, 3 regions, 4 LLMs)}. We are able to reduce GPU VM usage by 25\% through improved utilization, and VM cold-start by 80\% without violating SLAs. These translate into potential savings of US\$2.5M per month. 
\item Lastly, we place the workload trace data, scheduling logic and realistic simulator for validating them as open-source artifacts at \url{https://github.com/shashwatj07/SageServe}. These can serve as a testbed for the wider community.
\end{enumerate}
We also discuss related work (\S~\ref{sec:related}) and offer our lessons learned and conclusions (\S~\ref{sec:conclude}).

%% file: SC_25/SystemandApplication.tex
\section{System and Application Model}
\label{sec:sys-app-model}

\begin{figure*}[t]
\vspace{-0.1in}
    \centering
\includegraphics[width=.95\linewidth]{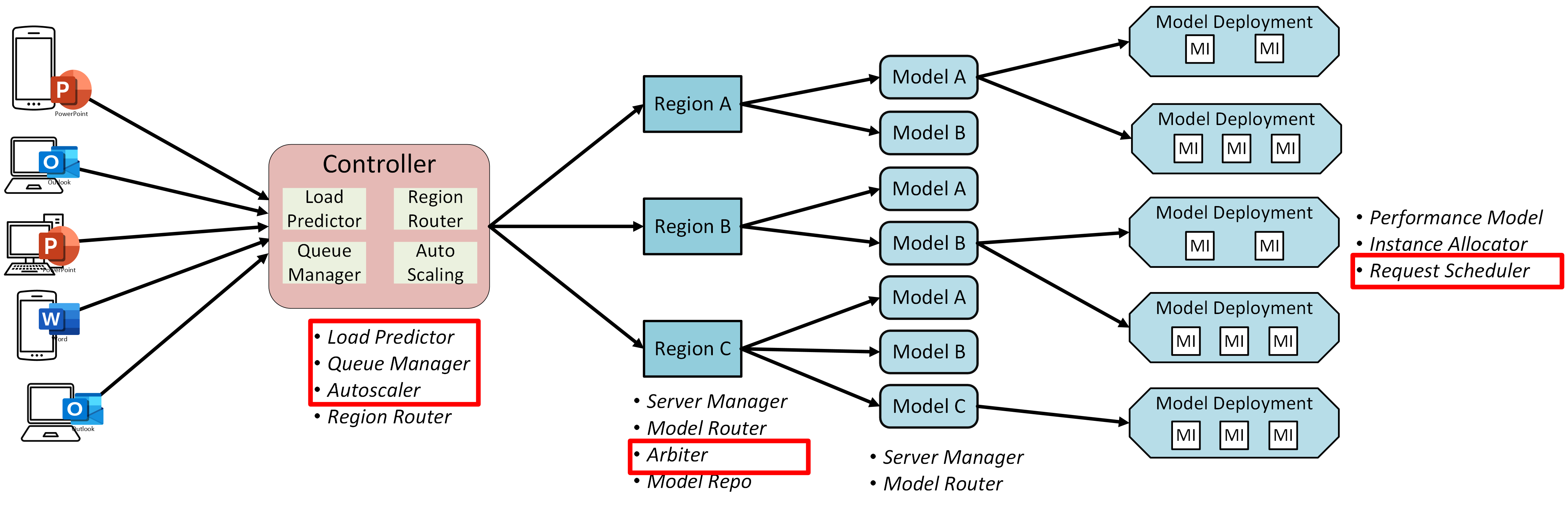}
\vspace{-0.15in}
    \caption{Overview of the \csp LLM serving architecture. The components added or improved upon (Load Predictor, Queue Manager, Autoscaler/Arbiter and Request Scheduler) by \sys are outlined.}
    \label{fig:simulator}
    \vspace{-0.1in}
\end{figure*}

We describe the cloud system, LLM deployment and workloads, motivated by global installations at \csp.

\subsection{Cloud VMs and LLM Model Instances}
The CSP's system model comprises multiple data centers (\textit{regions}) connected with high bandwidth network, with$\approx 50ms$ inter-region latency. Regions are assumed to be in USA (e.g., US-West, US-Central, etc.) to avoid issues of data sovereignty. Each region has thousands of \textit{GPU VMs}, e.g., Azure ND, with exclusive access to GPUs like NVIDIA A100/H100 or AMD MI300X and all its server resources. These can host LLM instances within regional capacity limits.

There are several standard LLM \textit{model types} that are available, e.g., Llama 2/3.3/4, GPT 3.5/4, Bloom, etc. with associated  default weights or custom weights based on fine-tuning (\autoref{fig:simulator}). A \textit{model instance} is one copy of a \textit{model type} that can serve requests. Each instance may require multiple GPUs depending on the size of the LLM, e.g., GPT3 may need 9 H100s while Llama-3 needs 4 H100s~\cite{mei2024helix}. Each VM is exclusively to one LLM instance. There can be multiple instances of an LLM type in a region as part of a deployment. A \textit{model endpoint} for that region receives requests to a model type and routes it to one of the instance deployment in a round robin manner. There can be constraints on the minimum and maximum instance counts per endpoint, for robustness and to avoid one model dominating.

\textit{Tokens per Second (TPS)}, the sum of input and output tokens processed per second forms the key throughput capacity metric, and we focus on input TPS
that a model instance can serve. The VM type and model type will determine the instance's performance, defined as input TPS achieved at a target latency~\cite{mei2024helix}. 
E.g. Llama2-70B~\cite{Llama} and Bloom-176B~\cite{bloom} models can achieve Q1--Q3 performance of 68--293 TPS and 50--177 TPS respectively on $8\times$ Nvidia A100 GPUs; this improves to 95--522 TPS and 82--397 TPS respectively with $8\times$ H100 GPUs. 

\subsection{LLM Workload Tiers and SLAs}
\csp supports multiple LLM inference workloads. \textit{Interactive Workloads \textbf{(IW)}} with \textit{low latency constraints (O(seconds))} from client-facing applications like chatbots, code generation, and email suggestions, require ``fast'' serving. Within these, some may require an even faster serving (IW-F) ($<1s$) while others may have a normal interactive latency goal (IW-N). 
In contrast, \textit{Non-Interactive Workloads \textbf{(NIW)}}, such as nightly document summarization on enterprise repositories or deep content generation, have \textit{relaxed deadlines (O(hours))}, and tolerate ``slow'' (or ``no'') serving. 

For IW tier, clients for each product or service, {e.g., Copilot for Word, Teams etc.,} may use one or more pre-defined LLM types with thousands of input/output tokens per request,
with tens of thousands of daily clients. 
For simplicity, we assume all clients are US-based since the regions are in the US.
NIW also uses a set of pre-defined LLMs whose architectures often overlap with IW but have a lower and non-periodic request rate that is stable through the week. We discuss these in \S~\ref{fig:wl}.

IW and NIW tiers have \textit{different SLAs} defined. \textit{Time to First Token (TTFT)} is the time between receiving a prompt to emitting the first response token and indicates responsiveness. In contrast, \textit{End-to-End (E2E) time} is the time taken for generating all output tokens for a request, and influences both latency and throughput. IW primarily have TTFT as SLA, typically $<1s$ for IW-F and $<1min$ for IW-N
at the $95^{th}$ percentile ($P95$), but can also have an E2E SLA. The SLA for NIW is typically a \textit{deadline} for \textit{batch completion} (e.g. $24h$ to summarize a document repository) with less emphasis on per-request latency. We assume that serving an IW request within its latency SLA accrues a \textit{utility} for the CSP, and serving an NIW request before its deadline has a \textit{(lower) utility}.

\subsection{Request Routing, Scaling and Scheduling Layers}
\para{Routing Mechanisms}
All IW requests are routed through a common LLM API service~\cite{AzureBatchAPI} to one of several LLM endpoints that can serve them (\autoref{fig:simulator}). This \textit{global routing} to one of the available regions is based on network latency, spatial proximity or the current load on the region's endpoints. Then, a \textit{region router} sends requests to 
endpoints for that model within that region, and further to instances within the selected deployment in a round robin manner to balance the load and address token skews. We assume a managed network and trusted security environment. There are no other security constraints that limit the mapping of instances to VM or requests to endpoints.

\para{Scaling Delays} 
Creating a new 
LLM instance on VMs when scaling up a model type in a region (\autoref{fig:arch}) has several \textit{provisioning costs} that can vary based on the conditions. Allocating VMs to the instance is an initial cost. If these VMs do not have the LLM already deployed on them, the model architecture and weights need to be copied and instantiated on them. This cold-start time depends on the model size, and on whether they are available in a local repository in that region (e.g., $\approx10mins$), or need to be moved from a remote region (e.g., $\approx 2h$). If the VMs already have the LLM architecture deployed from a prior provisioning but with different weights, only the weights need to be updated and the latency reduces. When an instance is being provisioned, the VMs are not available for use. So the model provisioning time constitutes \textit{wasted GPU cycles}. There are also latency costs to acquire a GPU VM and update all upstream services such as load balancer, etc. Since this total time takes mins--hours, frequent re-provisioning is inefficient.

\para{Donating to Spot Instances}
The workloads are executed on LLM instances that are provisioned in a private network space for O365. However, if the endpoints of common Azure LLM model types are idle, they can be leased to external users of Azure AI as (preemptible) \textit{spot LLM instances} for inferencing at a lower cost, and reclaimed when the internal demand increases. Switching an instance from a private to a spot role, and the reverse, is relatively fast, $\approx 1min$. Typically, the \textit{utility benefits} of leasing out spot instances is lower than that gained from executing the internal IW and NIW workloads. But this is still better than keeping the VMs idle. During some periods, $25\%$ of instances in a region may be donated to spot LLM instances; \textit{this is a lost opportunity cost we aim to fix} by re-routing and running NIW workloads on them.

\para{Scheduling Algorithms} Each LLM instance employs a scheduling policy that selects the next batch of requests from the waiting queue. The scheduler has access to deterministic request properties including prompt token count, service level agreement (SLA) tier, and arrival timestamp, utilizing these attributes alongside available GPU memory to make batching decisions. To minimize computational overhead from redundant processing, requests remain non-preemptible within a batch until memory exhaustion occurs on the virtual machine.

\section{Workload Characterization} \label{sec:wlstudy} 
In this section, we analyze traces for IW-F, IW-N and NIW workload tiers for O365 applications collected for 1 week each in July, 2025 and November, 2024. 
Specifically, we examine request traces from four popular OpenAI models deployed in three regions: US East, US West and US Central. 
To maintain anonymity, we refer to the models as Model A, B, C and D. The traces are made publicly available. Model A has a relatively larger number of parameters compared to Models B, C and D.
We focus on the following key dimensions: (1) workload demand from different tiers, characterized in terms of Requests Per Second (RPS) and Tokens Per second (TPS); (2) latency trends; and (3) capacity utilization trends. {TPS includes sum of input and output tokens per second.}

\begin{figure*}[t]
\vspace{-0.1in}
    \centering
    
\subfloat[1 Week in November 2024]{\includegraphics[width=0.35\textwidth]{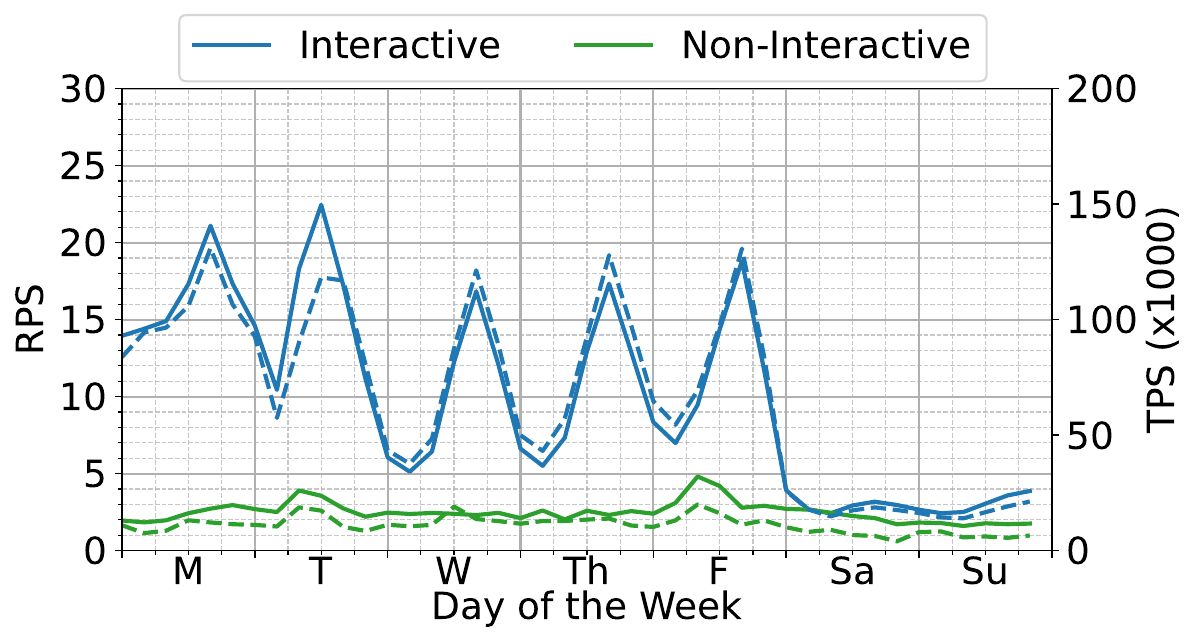}\label{fig:wl:allv0}}
\subfloat[Nov'24 1h]{\includegraphics[width=0.14\textwidth]{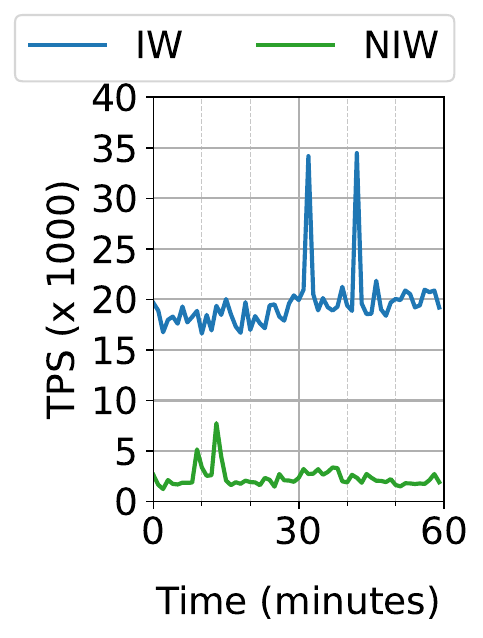}\label{fig:wl:all1h}}
\subfloat[1 Week in July 2025]{\includegraphics[width=0.35\textwidth]{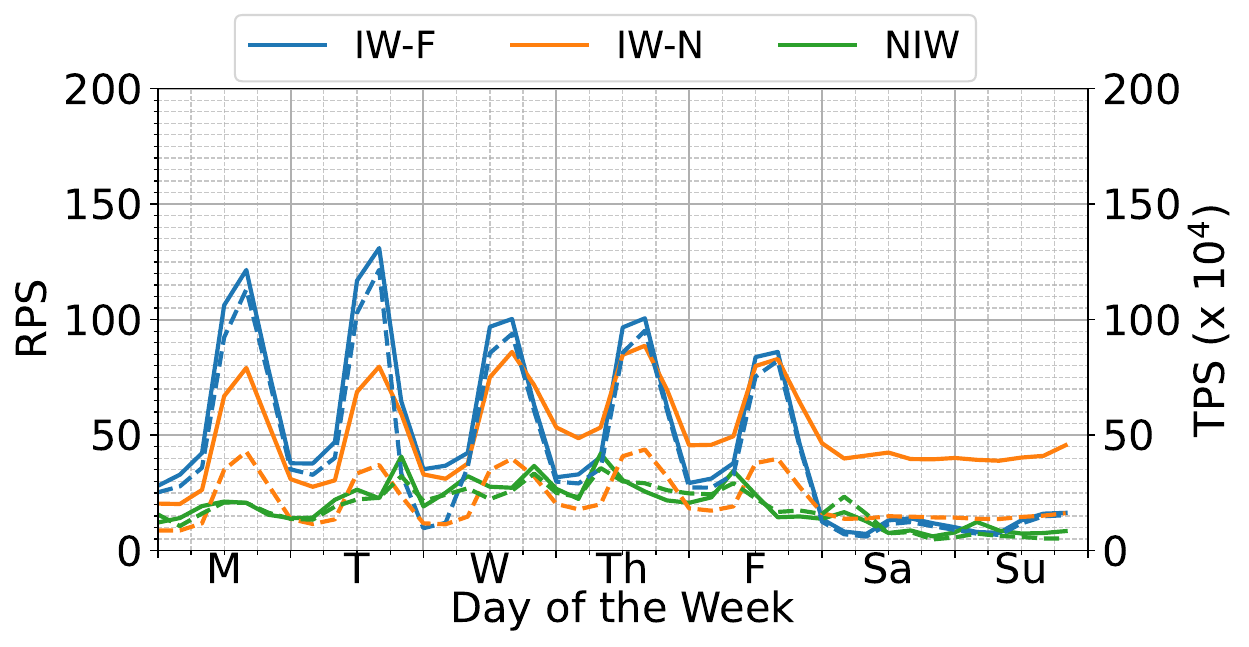}\label{fig:w2:cumload}}
\subfloat[Jul'25 1h]{\includegraphics[width=0.14\textwidth]{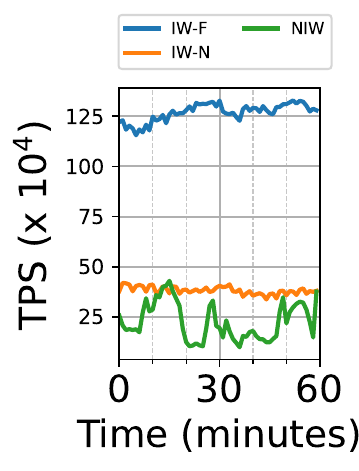}\label{fig:w6:1hload}}
\vspace{-0.1in}
\caption{Aggregated RPS \textit{(solid line)} and Total Input+Output TPS \textit{(dashed line)} for IW \& NIW for 3 US regions.}
    \label{fig:wlcummu}
    \vspace{-0.1in}
\end{figure*}

\begin{figure*}[t]
    \centering
\subfloat[\textbf{IW-F:} West US  
]{\includegraphics[width=0.33\textwidth]{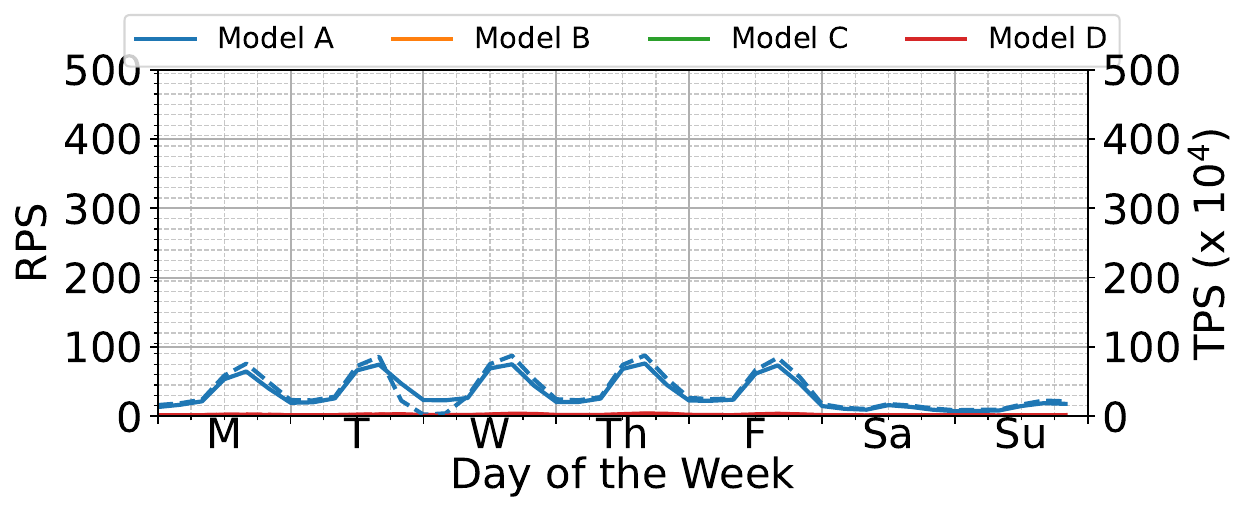}\label{fig:w2:west}}%
\subfloat[\textbf{IW-F:} Central US region]{\includegraphics[width=0.33\textwidth]{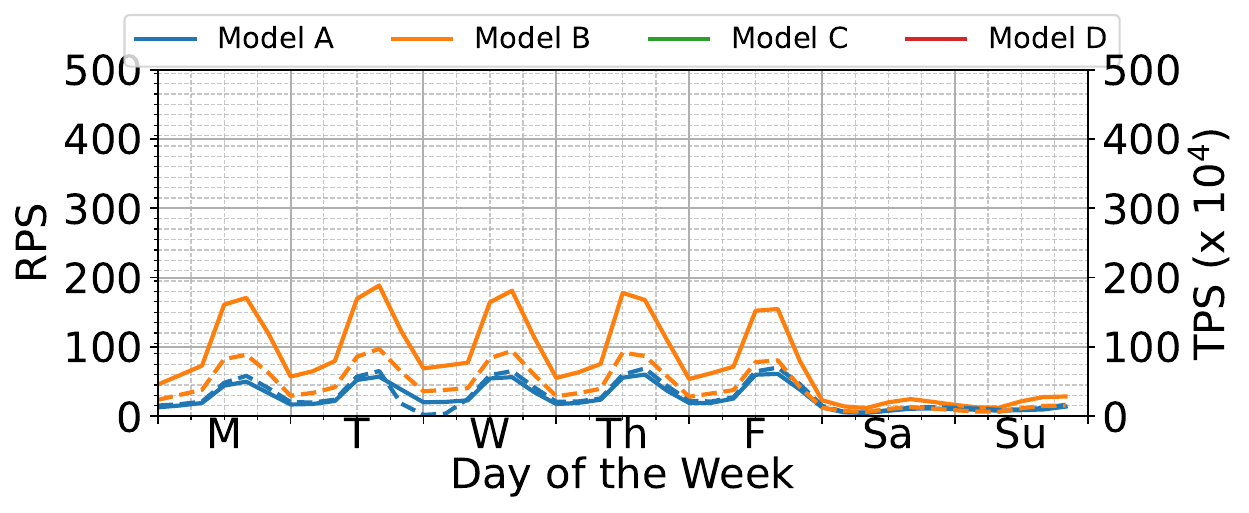}\label{fig:w2:central}}%
\subfloat[\textbf{IW-F:} East US region]{\includegraphics[width=0.33\textwidth]{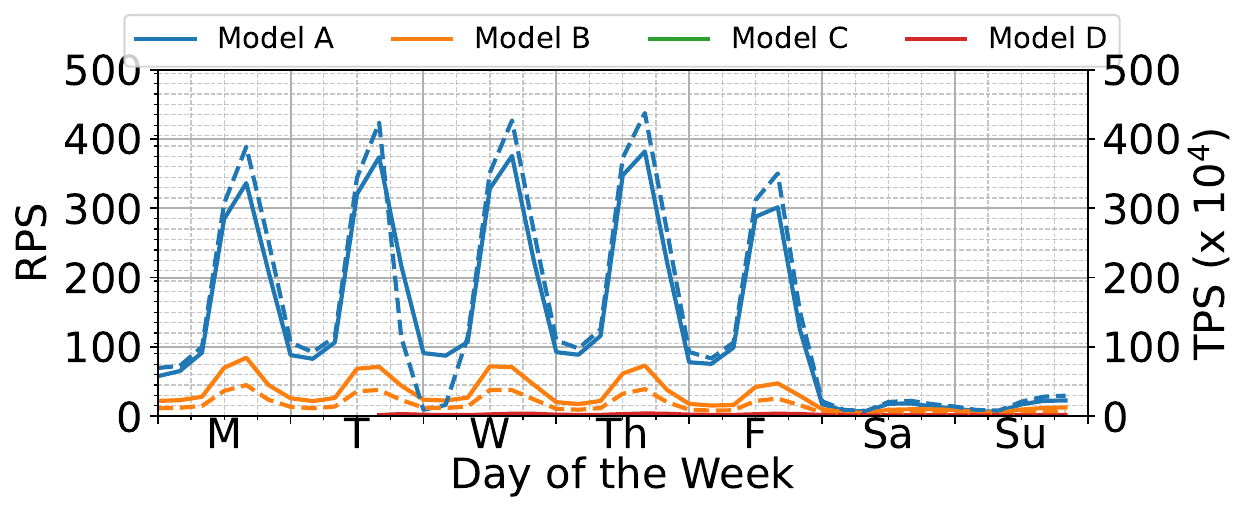}\label{fig:w2:east}}
\\
\vspace{-0.1in}
\subfloat[\textbf{IW-N:} West US region]{\includegraphics[width=0.33\textwidth]{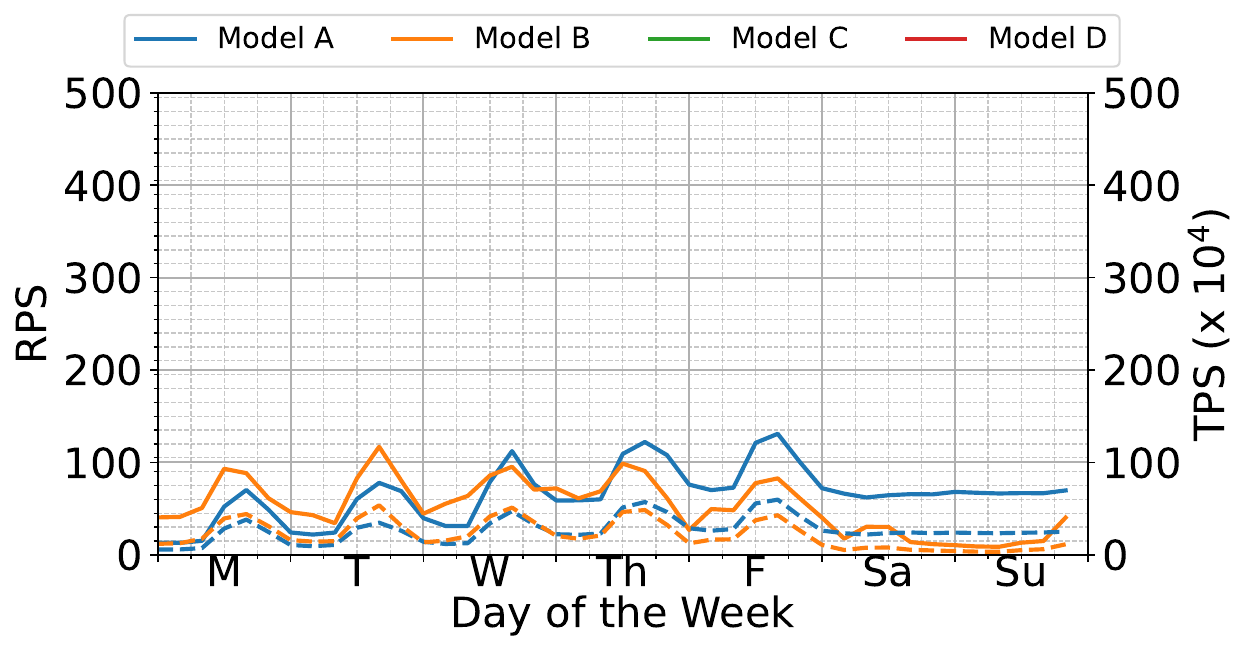}\label{fig:w3:west}}%
\subfloat[\textbf{IW-N:} Central US region]{\includegraphics[width=0.33\textwidth]{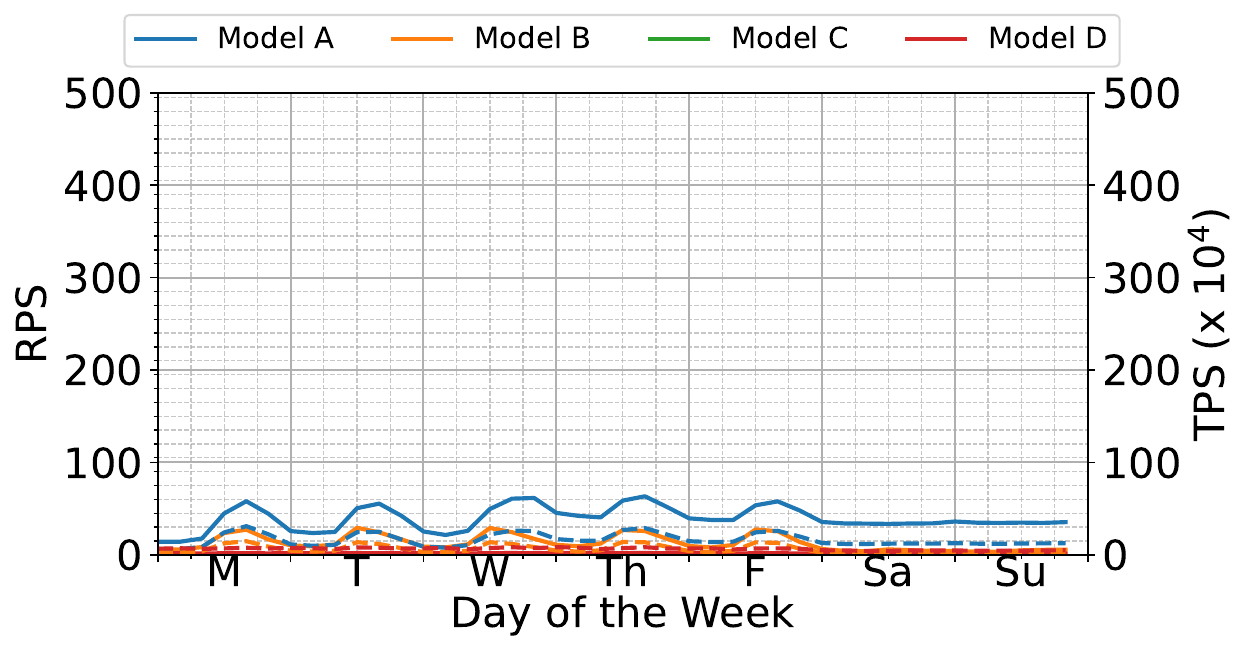}\label{fig:w3:central}}%
\subfloat[\textbf{IW-N:} East US region]{\includegraphics[width=0.33\textwidth]{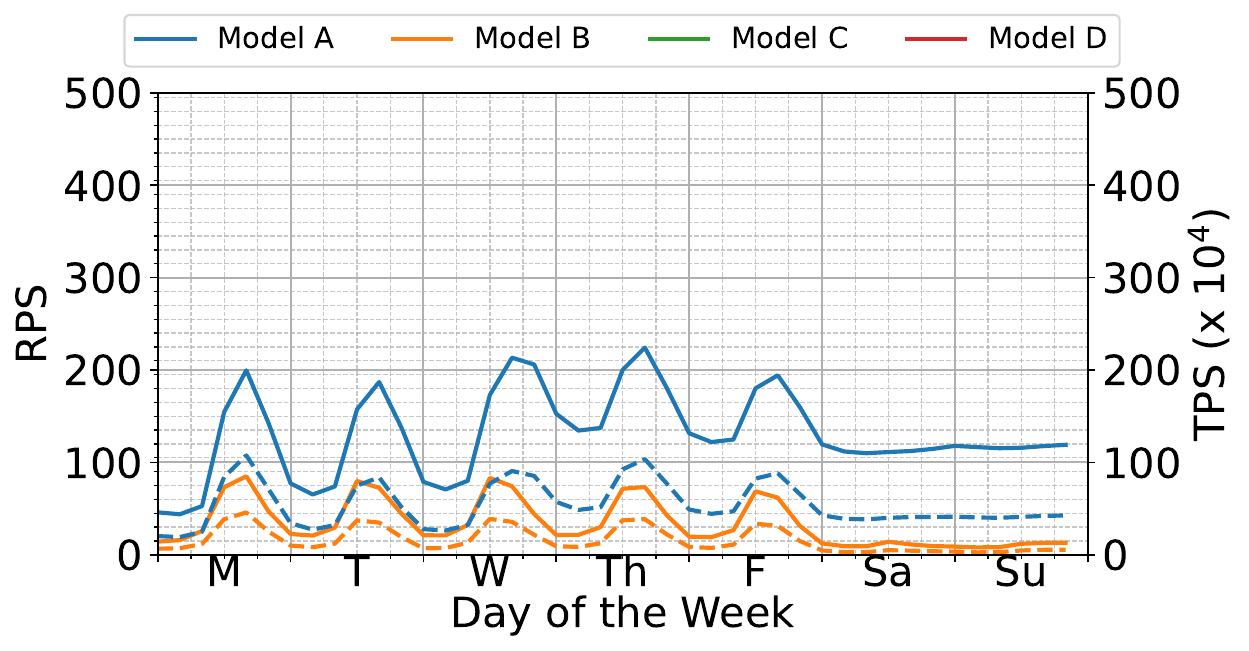}\label{fig:w3:east}}
\\
\vspace{-0.1in}
\subfloat[\textbf{NIW:} West US region]{\includegraphics[width=0.33\textwidth]{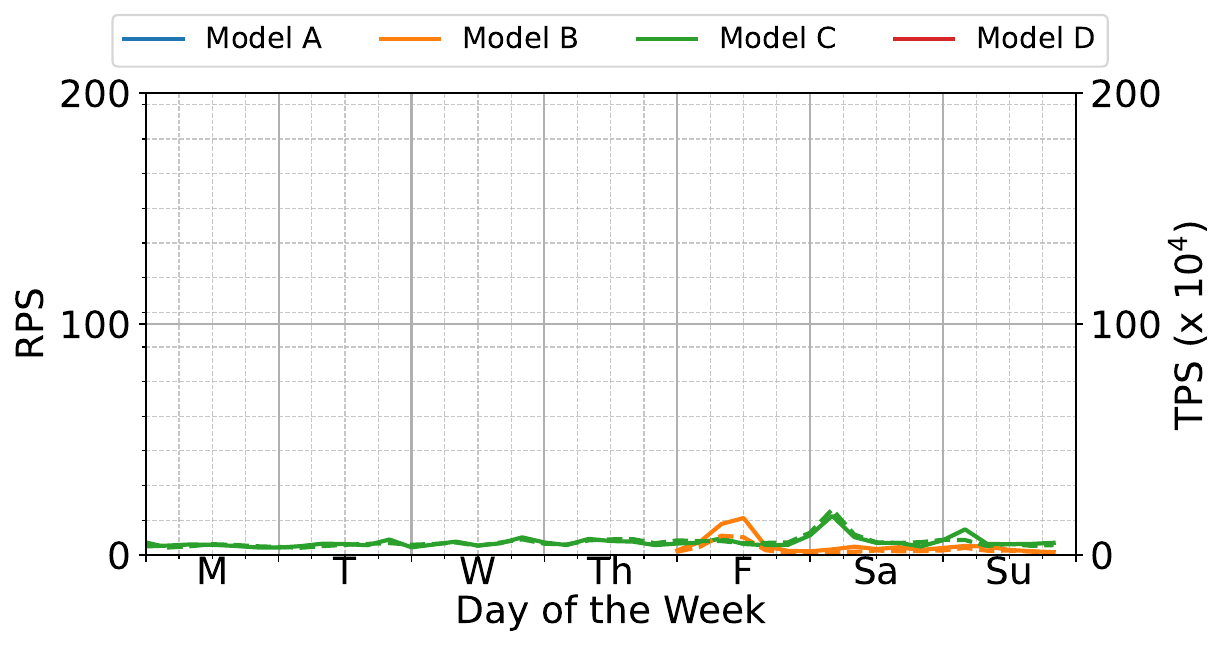}\label{fig:w4:west}}%
\subfloat[\textbf{NIW:} Central US region]{\includegraphics[width=0.33\textwidth]{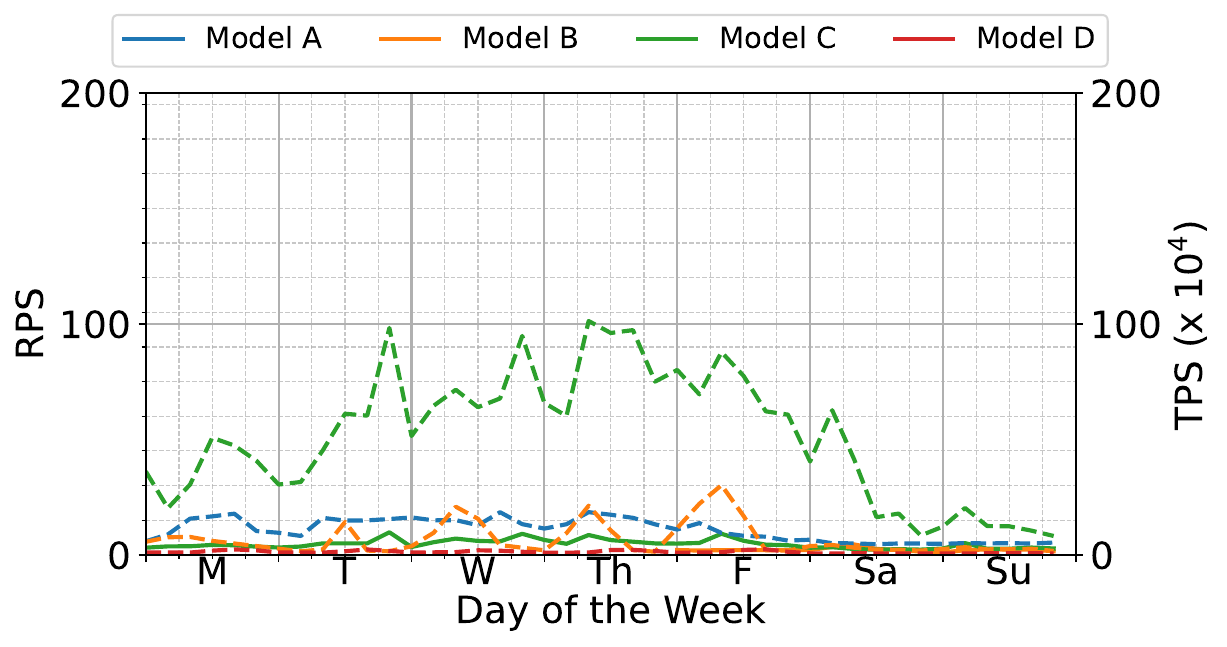}\label{fig:w4:central}}%
\subfloat[\textbf{NIW:} East US region]{\includegraphics[width=0.33\textwidth]{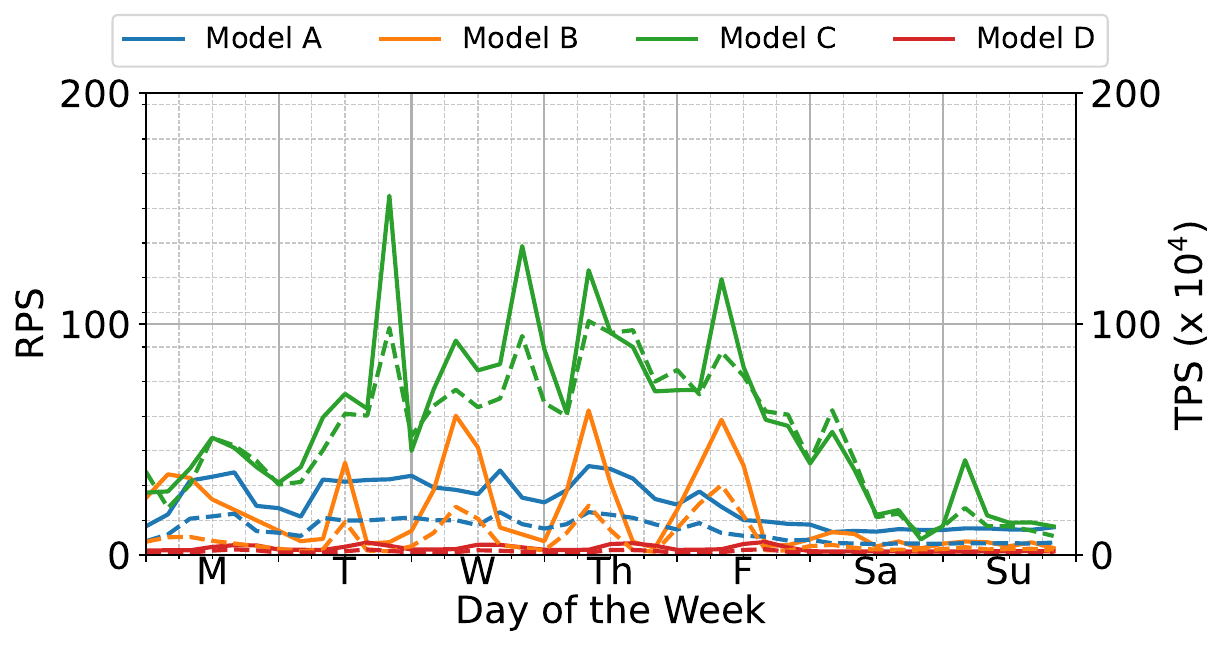}\label{fig:w4:east}}\\
\vspace{-0.1in}
    \caption{RPS \textit{(solid line)} and Total Input+Output TPS \textit{(dashed line)}  per model for 1 week in \textit{July 2025}, for IW-F, IW-N, and NIW requests (rows) in three US regions (columns).}
    \label{fig:wlnew}
    \vspace{-0.1in}
\end{figure*}

\begin{figure*}[t]
    \centering
\subfloat[West US region]{\includegraphics[width=0.33\textwidth]{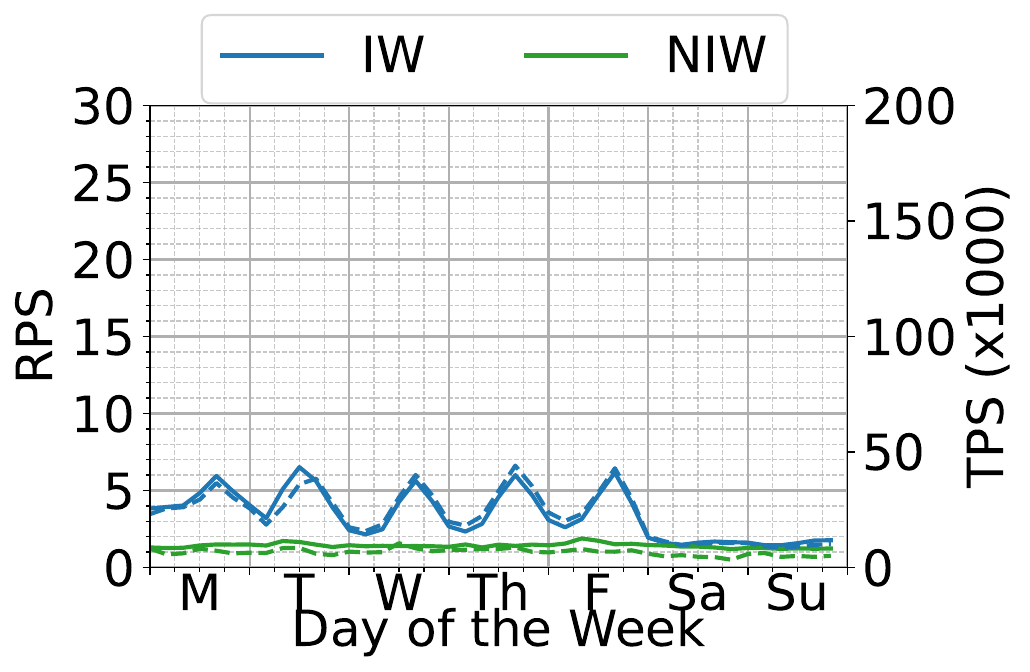}\label{fig:wl:west}}%
\subfloat[Central US region]{\includegraphics[width=0.33\textwidth]{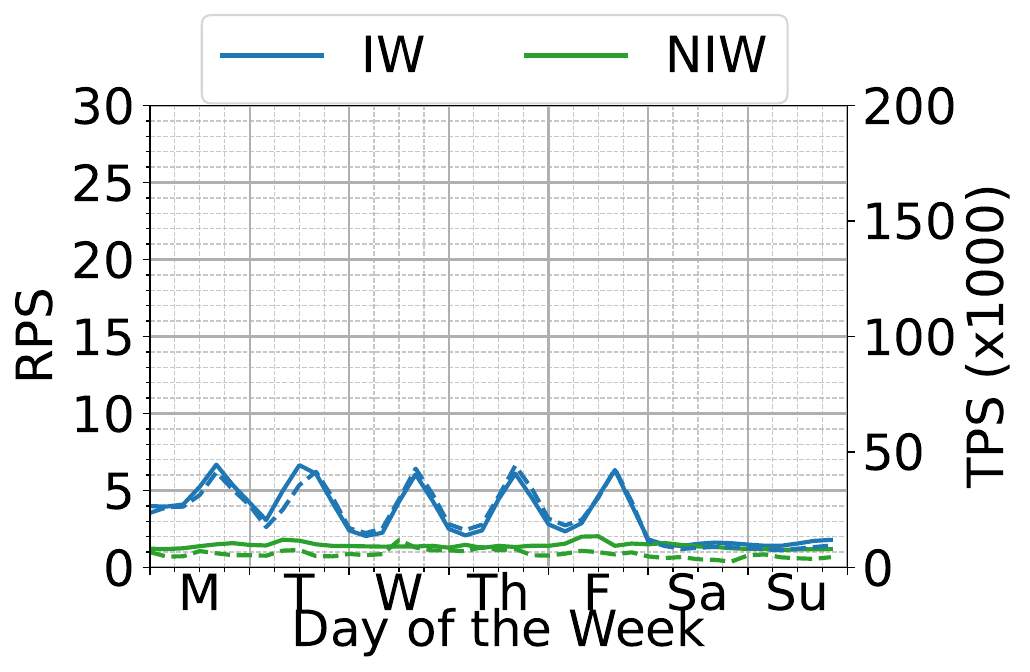}\label{fig:wl:central}}%
\subfloat[East US region]{\includegraphics[width=0.33\textwidth]{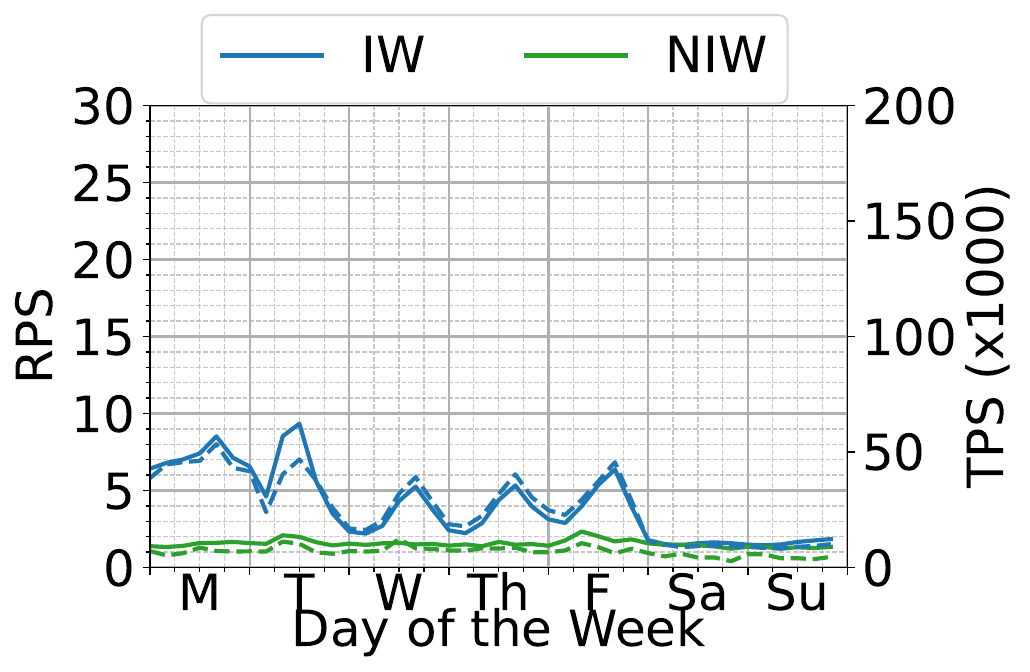}\label{fig:wl:east}}%
\vspace{-0.1in}
    \caption{RPS \textit{(solid line)} and Total Input+Output TPS \textit{(dashed line)} of IW and NIW requests, summed across 4 LLM models for 1 week in \textit{Nov, 2024} for three regions (a)--(c). 
    }
    \label{fig:wl}
    \vspace{-0.1in}
\end{figure*}

\para{Demand Patterns}
\autoref{fig:wl:allv0} and \autoref{fig:w2:cumload} illustrate the cumulative load (RPS and TPS) from the 4 models and 3 regions, for each of the 3 tiers. The \textit{Nov 2024 trace} does not distinguish between IW-F and IW-N as that is a recently introduced feature, reflecting the evolving nature of LLM-based offerings.
\textit{IW-F workloads} exhibit clear diurnal periodicity with weekends quiescing, indicating a high degree of predictability. The periodicity score ($ps$) of Model A varies from  $0.7$--$0.95$, indicating a strong positive autocorrelation between daily traffic patterns. 
\textit{IW-N traces} also show periodicity, with trends similar to IW-F, having $ps \approx 0.6$--$0.8$ for Model A and $0.3$--$0.5$ for Model B. In contrast, \textit{NIW traces} are less predictable ($ps=0$--$0.286$), without clear periodicity. Notably, IW-F form the largest fraction of all requests and TPS, followed by IW-N, which together are $72\%$ of all requests.

Next, we analyze the properties of requests from the different models in each tier and region, as shown in \autoref{fig:w2:west}–\autoref{fig:w4:east} for the \textit{Jul, 2025 trace}. 
As observed from the cumulative load in \autoref{fig:w2:cumload}, \textit{IW-F requests} exhibit this on a per-region basis too and for all models. However, the amplitude of the traffic patterns vary by region, with East US exhibiting a high demand, followed by Central US and West US. 
{Since requests from any US client can be routed to any of the US data centers by the Region Router (\autoref{fig:simulator}), based on the load and latency, this does not necessarily reflect the client demand from those regions. For the same reason, we do not see any timezone phase shifts in load across the three regions.} {Inter-region routing is enabled by low network latency between regions and makes the problem more challenging due to the need for global optimization.} 
\textit{IW-N requests} also exhibit periodicity across most regions (\autoref{fig:w3:west}–\autoref{fig:w3:east}). However, in some cases, it shows growth in usage across different weekdays, e.g., Model B has more requests on Wed/Thu/Fri than Mon/Tue, visible in East and West US.
In general, model demand for IW workloads varies by region, while maintaining the overall trend. E.g., Model A (blue lines) is most popular in  East US, at $4\times$ the load as West US for IW-F, whereas Model B (orange) sees the highest demand in Central US (for IW-F) and West US (IW-N). This could be due to the imbalance in the capacity allocated to each LLM type per region. 
Consequently, the capacity required for each model differs across regions.

In contrast, \textit{NIW requests} are less predictable on TPS and RPS, and the request demand is lower in all regions, with West US having negligible requests. The popularity of NIW models also vary across regions. Interestingly, the TPS per request for Model C in Central US is much higher compared to other workloads due to the presence of a feature evaluation and testing application. This is due to the presence of batch-oriented workloads in this tier, which perform bulk operations.

\colortakeaway{\textbf{Summary}.
Latency-sensitive IW requests exhibit strong diurnal and weekday-versus-weekend periodicity within a region for each model. This pattern should be leveraged to design systems that optimize resource allocation for anticipated demand.
The lower load, poor periodicity but flexible SLA for NIW makes them suitable for opportunistic, deferred execution.
} 

\begin{figure*}[t]
\vspace{-0.1in}
    \centering
      \subfloat[Top applications]{\includegraphics[width=0.35\textwidth]{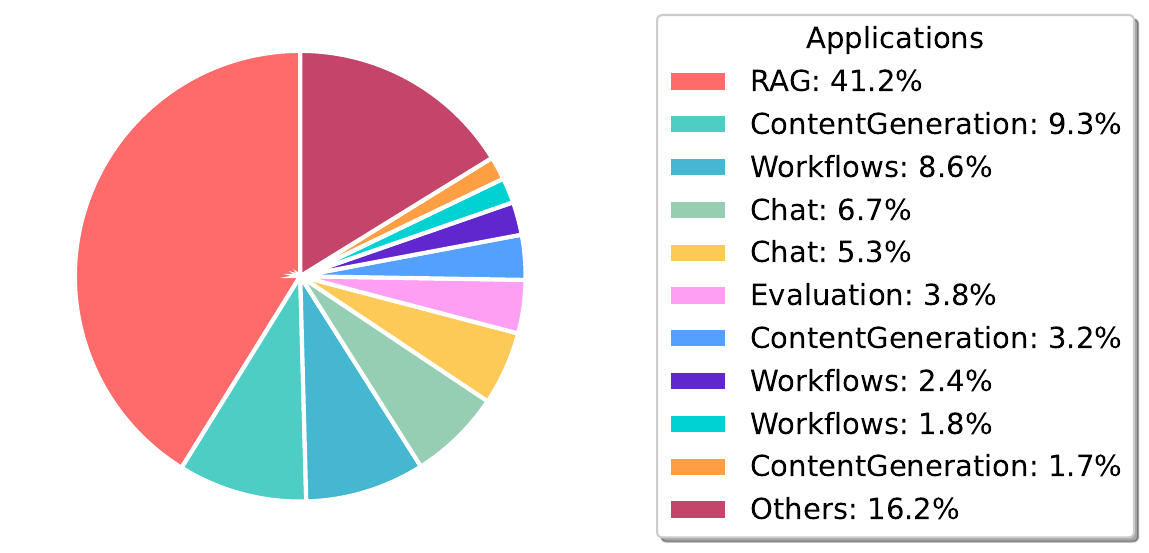}\label{fig:w6:pie_top}}%
    \subfloat[RPS \& TPS trends of Top Apps]{\includegraphics[width=0.35\textwidth]{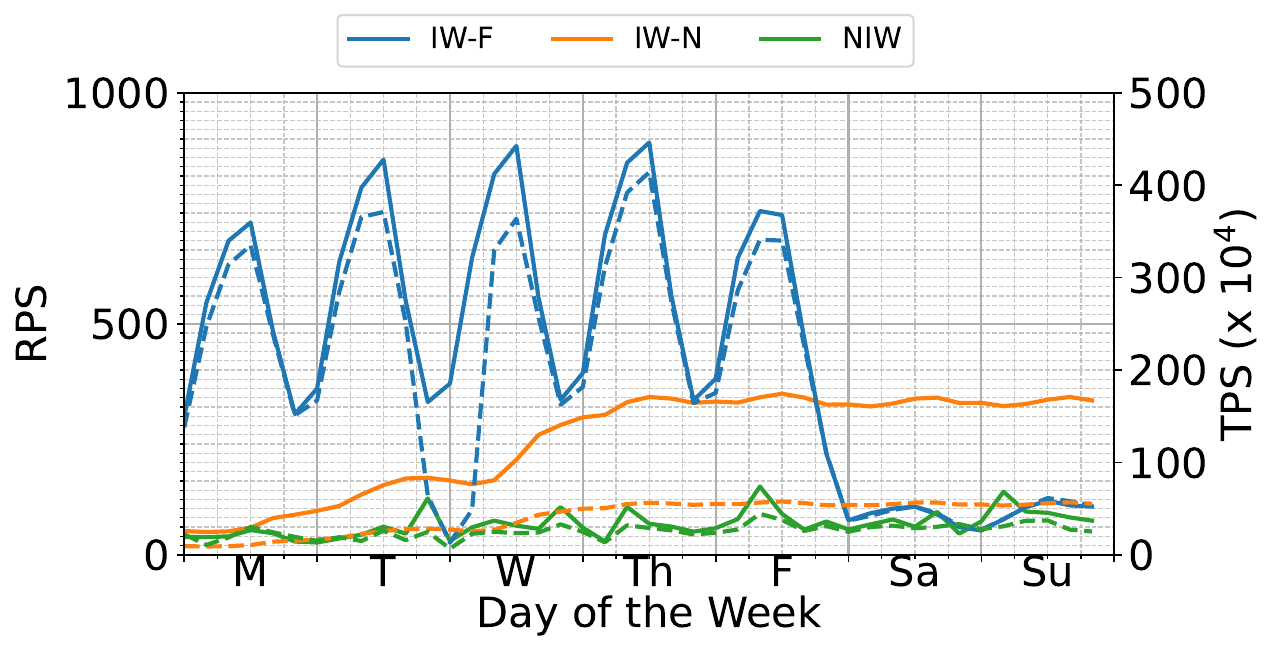}\label{fig:w6:rpstps_top}}%
\subfloat[Latency Plot (One day trace)]{\includegraphics[width=0.3\textwidth]
{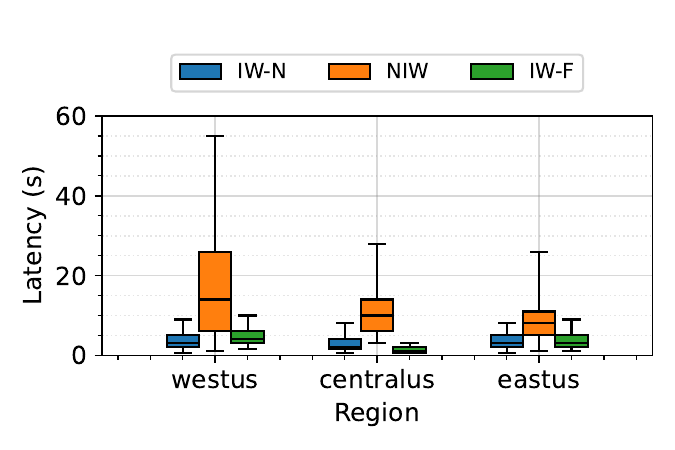}\label{fig:w6:latspike}}\\
\vspace{-0.1in}
\subfloat[Latency plots from peak 1h]{\includegraphics[width=0.43\textwidth]{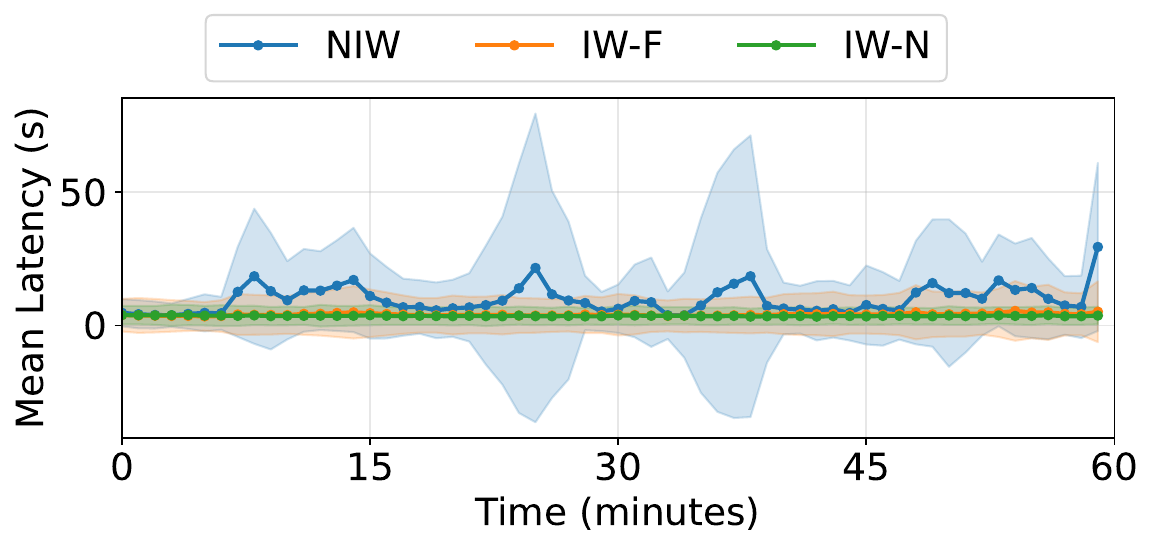}\label{fig:w6:peakhourlat}}%
\qquad
\subfloat[Normalized IW-F TPM for Model A]{\includegraphics[width=0.39\textwidth]{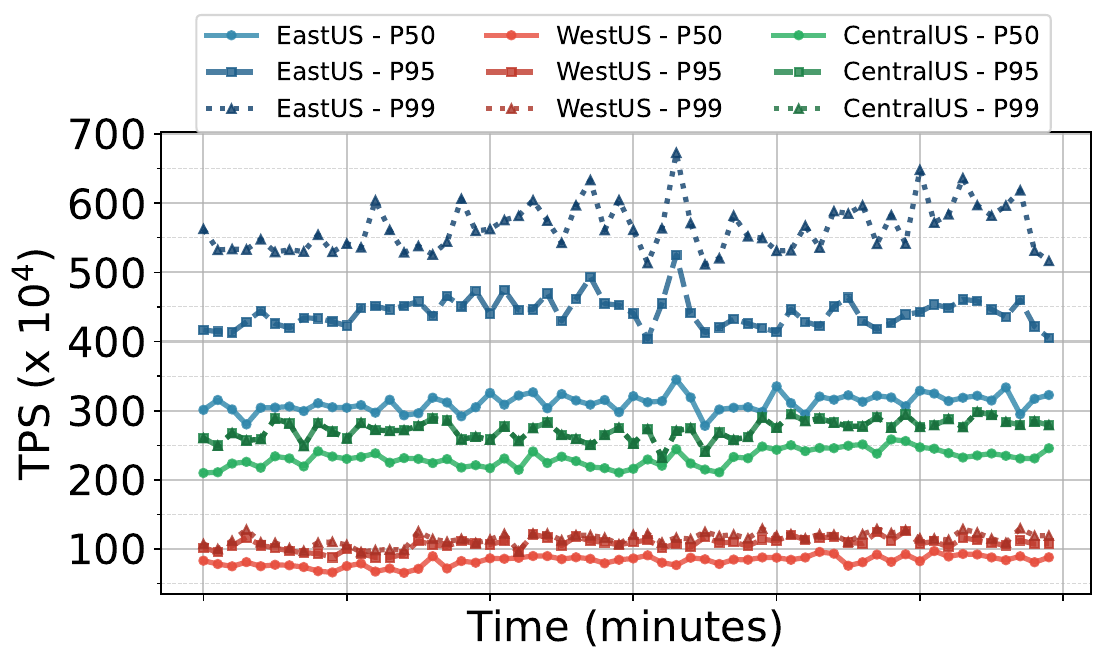}\label{fig:w6:1hcapacity}}
\vspace{-0.1in}
    \caption{(a) Top applications across different workloads. (b) RPS and TPS of top applications over 24h. (c) E2E request latency for workloads and regions. (d) Latency variation for the peak 1h of the cumulative. (e) Trans. per Min. (TPM) at at different percentiles from among model deployments within each region for 1h, indicating diversity of load. All for \textit{July 2025} trace.
    }
    \label{fig:wfg}
    \vspace{-0.1in}
\end{figure*}

\para{Evolution of Workloads Over Time}
\autoref{fig:wl:allv0} and \autoref{fig:w2:cumload} compare the shift in cumulative workload trends across the two traces 7-months apart, and similarly \autoref{fig:wlnew} and \autoref{fig:wl} for each region. Both the request rate and TPS have increased during this time by $\approx 5\times$. New workloads include conversational agents, RAG scenarios, and 
feature testing and evaluation frameworks, driven by the increasing adoption of Copilot in O365 applications and the emergence of new LLM workloads and applications. This happens across regions as well, while still retaining the periodicity. This underscores the critical need for improved capital efficiency of GPU resources in the LLM serving design, all the more so given the long lead times for acquiring new hardware and the need to conserve operational costs while delivering SLA.
In addition, the workload tiers have also evolved during this period with the introduction of fast and normal latency tiers within the interactive workload. This can further evolve into a continuum from fast to slow, and high to low priority workloads over time. 

\colortakeaway{\textbf{Summary}. As new workloads to emerge with varying latency needs and priorities, the serving system must scale efficiently -- without leading to a disproportionate increase in capacity needs.}

\para{Application-level Workload Patterns}
\autoref{fig:w6:pie_top} shows the generic names of the \textit{top 10 applications} in O365 leveraging LLMs, based on the request count from the day of the week with the highest traffic, {cumulatively across all tiers}. Notably, $41.2\%$ of the requests originate from a Retrieval-augmented Generation (RAG) system, which explains the high number of tokens processed per second. Other top applications include context-driven use cases such as insights generation, content creation, chat applications, and evaluation frameworks.
\autoref{fig:w6:rpstps_top} reports the RPS and TPS trends for just these top application across different latency tiers. As seen earlier, latency-sensitive IW-F requests continue to have a diurnal pattern while IW-N shows a gradual growth, both of which are predictable using simple forecasting models we later discuss. Additionally, NIW requests maintain a consistent load throughout the week. The mean TPS for the  NIW workloads is $\approx 177.5 \times 10^4$ TPS, indicating a relatively stable load throughout  the observed time period.

While the daily trends across 10s of minutes or hours exhibit patterns, this does not carry forth at \textit{fine-grained time scales}.
\autoref{fig:w6:latspike} shows the latency distribution across various tiers for the three regions, {aggregated across the top applications}. NIW workloads exhibit greater variance in latency, as expected, since they have a relaxed SLA with mean latency of $15s$ and variance of $120s^2$. The variance ratio of NIW:IWF:IW-N is $150:1:0.5$. The E2E latency trends vary by region, with IW-F showing region-specific variability that indicates load imbalance. E.g.,  West US has (mean, median,  $P95$ latency) of $(4455 ms,\; 2700 ms,\; 14,539 ms)$, while Central US shows
 $(3349 ms,\; 618 ms,\; 14,967 ms)$, and East US
$(3615 ms,\; 1840 ms,\; 11,179 ms)$. 
We zoom in and analyze the load (\autoref{fig:w6:1hload}, \autoref{fig:wl:all1h}) and latency (\autoref{fig:w6:peakhourlat}) during the peak hour of the day. These plots reveal noticeable latency spikes at short time scales of $1min$, highlighting greater variability in IW-F latency compared to IW-N. E.g., the blue area plot in \autoref{fig:w6:peakhourlat} shows the {standard deviation} in the latency for IW-F, which changes a lot indicating potential load imbalance.
This imbalance is also seen from the load served by model deployments within a region (\autoref{fig:w6:1hcapacity}), where $P50$, $P95$ and $P99$ load are reported across time per region for {Model A}. While West US (orange) and Central US (green) serve similar loads using different instances, East US (blue) exhibits divergence across the $P50$--$P99$ latencies, indicating uneven load distribution across model instances in that region. This imbalance is further corroborated by the fact that East US accounts for $52\%$ of the cumulative capacity allocated to Model A in the three regions, followed by West US ($27\%$) and Central US ($21\%$). These highlight the need for fine-grained resource scheduling within regions. 

\section{Approach to Effective Resource Scaling} \label{sec:empstudy}

\begin{figure}[t]
\vspace{-0.1in}
    \centering
\subfloat[\textit{Siloed.} Separate instance pool for IW/NIW requests leads to under utilization of GPUs.]{
\qquad
\includegraphics[width=0.22\textwidth]{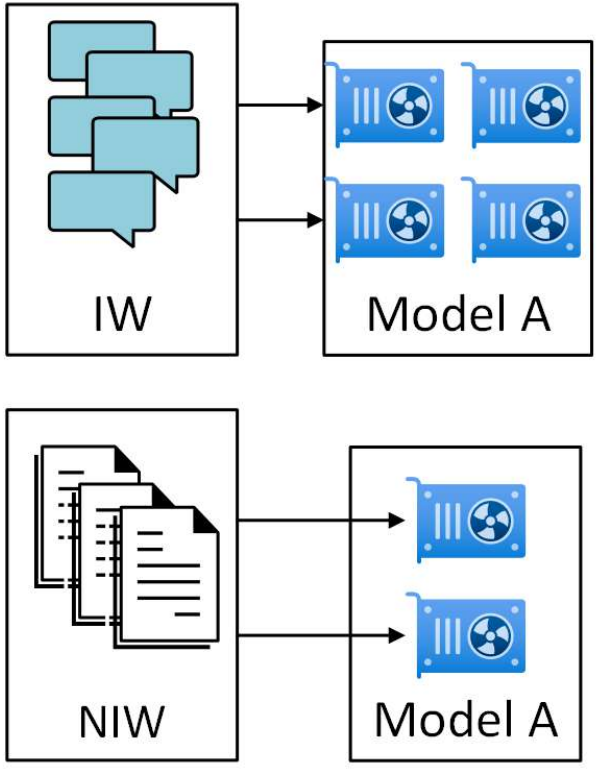}\label{fig:arch_silo}
\quad
}~~
\subfloat[\textit{Unified Reactive Scaling.} New instances are added when util. exceeds threshold, causing SLA violation.]{
\quad\includegraphics[width=0.28\textwidth]{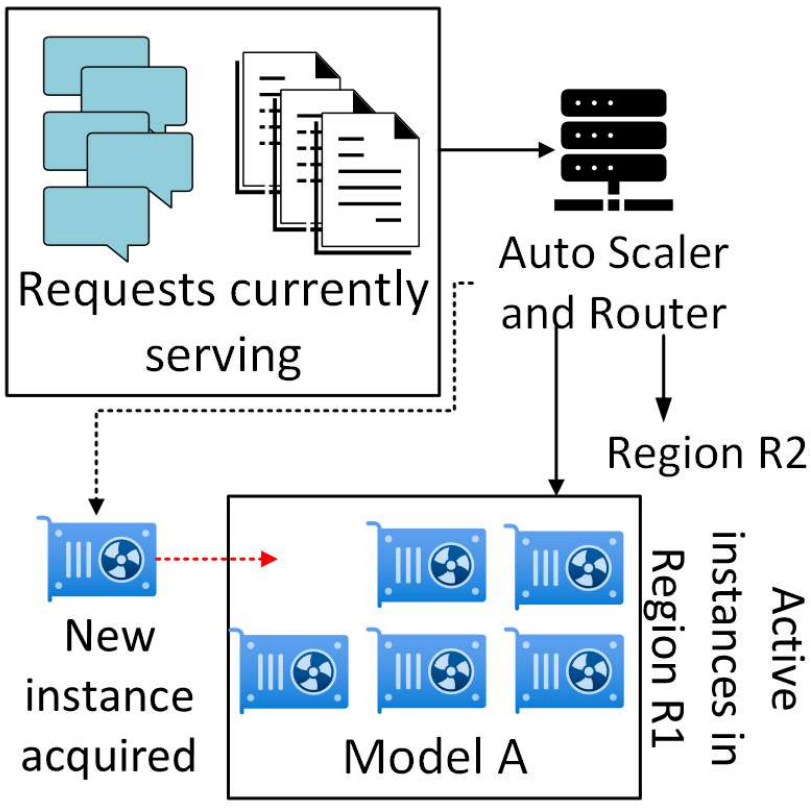}\label{fig:arch_react}
\quad
}~~
\subfloat[\textit{Predictive Scaling.} Additional instances required for IW are allocated \textit{a priori}, based on forecasts.]{
\quad
\includegraphics[width=0.28\textwidth]{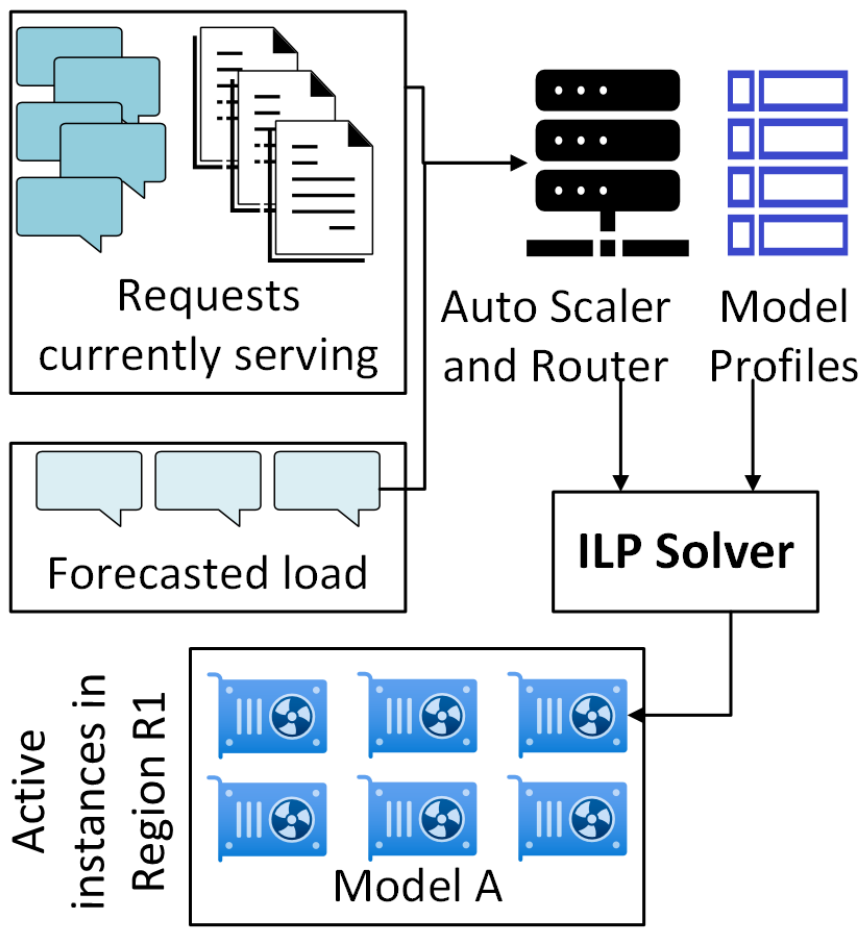}\label{fig:arch_predict}
}
    \vspace{-0.05in}
    \caption{Routing and Scaling Strategies for LLM Inferencing}
    \label{fig:arch}
    \vspace{-0.05in}
\end{figure}

We study the effect of scaling resources allocated to an LLM type on GPU capital efficiency and SLA compliance for both IW and NIW workloads. Since GPU VMs are costly, the goal is to maximize their utility -- either by serving IW or NIW tiers, or leasing the idle capacity as spot instances.
The baseline \textit{siloed deployment} is currently used in \csp for serving LLM inference requests (\autoref{fig:arch_silo}). It maintains separate instance pools per workload (IW/NIW) and LLM type in each region. 
It applies a greedy scaling policy, adding instances to the region endpoint when the effective memory utilization for the instances increases $>70\%$, and returns an instance to spot pool if the utilization drops $<30\%$. The effective memory utilization excludes memory used for model weights, and is a reliable proxy for the request load as it essentially captures the KV Cache footprint of in-flight requests. These scaling decisions are made per request, constrained by instance limits per model type. However, siloing into pools fragments VMs, often leading to suboptimal utilization.

As an alternative, we propose a \textit{reactive scaling heuristic} to serve both IW and NIW tiers from a \textit{unified resource pool} for a model in a region (\autoref{fig:arch_react}), allowing dynamic sharing of spare capacity across workload types~\cite{chiron}. Further, NIW requests are queued and served only when instance utilization drops below a limit (e.g, $50$--$60\%$) or when the the SLA deadline for the request is approaching. This ensures that they leverage spare capacity without affecting IW performance or triggering a instance scale-up. This enables shared use of model instances across workload tiers and
improves overall VM utilization, rather than donate to spot instances.
Switching an LLM instance between spot and internal endpoints takes $1$~min, while changing the model hosted on a VM requires $\approx10$~mins. We trigger scaling based on the effective utilization seen at regional endpoints, with a $15$-second cooldown enforced between events.

We illustrate the \textit{siloed} and \textit{unified reactive} scaling heuristics for $4$ open-source LLMs: Bloom, Llama 2, Llama 3.1, Llama 3.2, deployed in the $3$ US regions. Each region has $20$ instances per model: $16$ for IW and $4$ for NIW in the siloed approach, and all $20$ in a unified pool with the unified heuristic. Here, we do not separate out IW-F and IW-N. The minimum and maximum instance counts per endpoint are $2$ and $3$, which is the current practice in O365, ensuring fault tolerance and load balancing.
We use a realistic simulation harness built on Splitwise~\cite{patel2024splitwise}, whose results closely match real-world behavior (see \S~\ref{sec:results:simulator}). 
We use $8\times$ A100s per instance, and replay $1$ day of workload requests from West US (Tue of Nov, 2024 in \autoref{fig:wl}), which has $1.4M$ IW and $0.2M$ NIW requests.

\begin{figure*}[t]
\vspace{-0.1in}
    \centering
    \subfloat[Model instance count and instance-hours]{\includegraphics[width=0.47\textwidth]{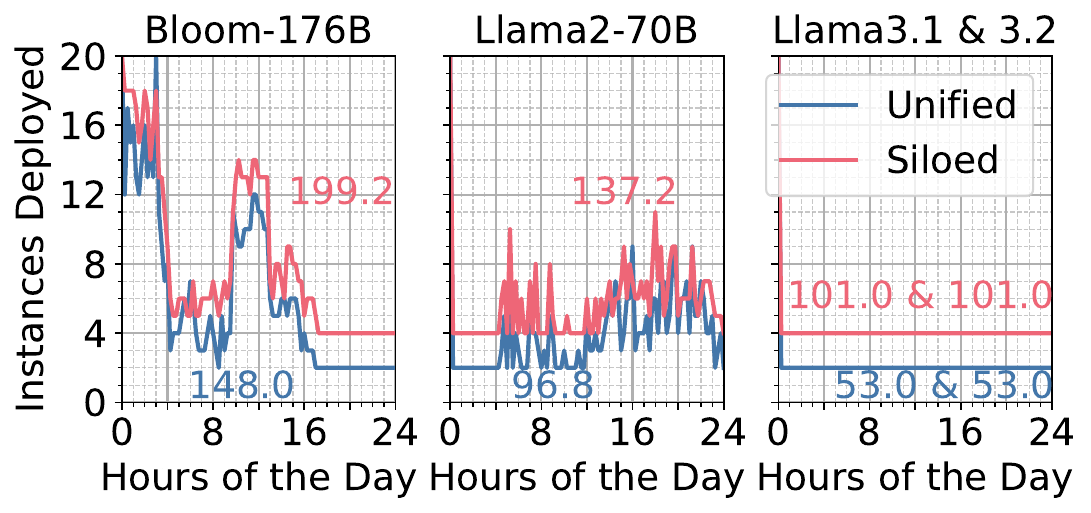}\label{fig:motivation-instance-hours}}%
\hfill
    \subfloat[Memory Utilization]{\includegraphics[width=0.45\textwidth]{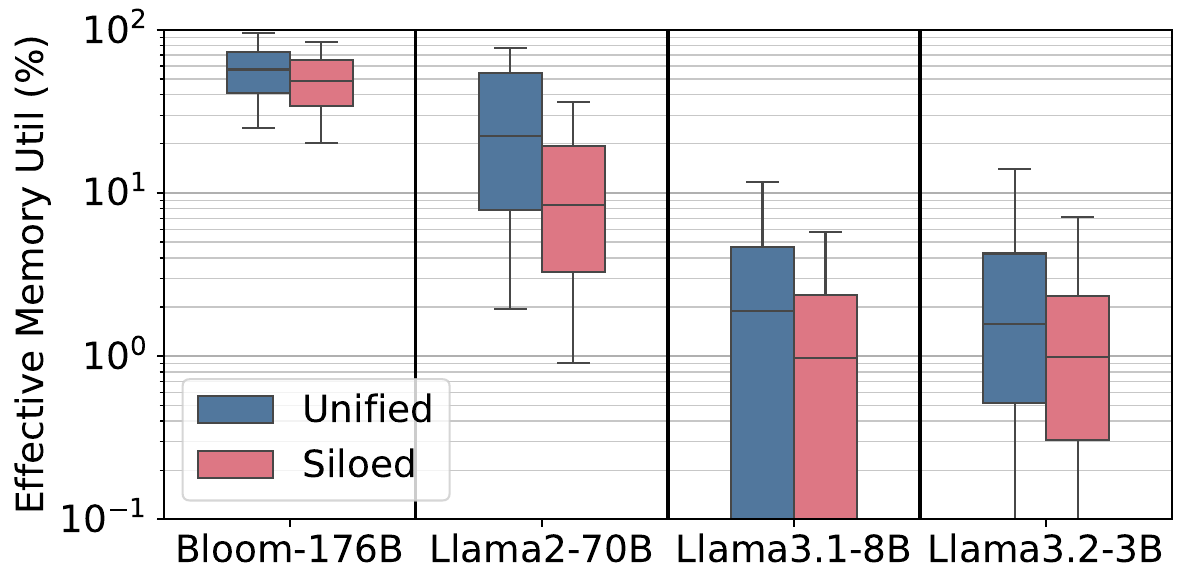}\label{fig:motivation-utilization}}%
    \vspace{-0.1in}
    \caption{Performance of Unified vs. Siloed strategies for workload trace in West US from Tuesday of Nov, 2024 (\autoref{fig:wl:west}), with peak 20 instances per model. Unified uses 34.5\% fewer instance-hours than the Siloed.}
    \label{fig:motivation-mem-util-combined}
\end{figure*}

\autoref{fig:motivation-instance-hours} shows the instance count at West US every 15-mins for the Siloed and Unified approaches for each model for that day. 
Our Unified deployment strategy consumes fewer model instance hours (area under the curve, shown as text label) for all four models despite using similar scaling thresholds for both deployments. This is because the instances in Unified are not tied to a workload type (IW or NIW) and can be used dynamically by either, based on demand.
Since Llama 3.1 and 3.2, being lighter in size, process tokens much faster, they maintained the minimum instance count throughout. Siloed allocates $2$ instances each for IW and NIW while Unified shares the same $2$ instances among IW and NIW requests.

\begin{table}[t]
    \centering
    \small
    \caption{95\%ile of TTFT and E2E latencies for serving different models using siloed and unified approach. 
    }
    \vspace{-0.1in}
    \label{table:silo_vs_reactive}
    \setlength{\tabcolsep}{2pt} 
    \begin{tabular}{l||r r|r r|r r|r r}
    \hline
        \bf Metric & \multicolumn{2}{c|}{\textbf{\textbf{Bloom-176B}}} & \multicolumn{2}{c|}{\textbf{\textbf{Llama2-70B}}} & \multicolumn{2}{c|}{\textbf{\textbf{Llama3.1-8B}}} & \multicolumn{2}{c}{\textbf{\textbf{Llama3.2-3B}}} \\ \cline{2-9}
        ~& \it Siloed & \it Unified & \it Siloed & \it Unified & \it Siloed & \it Unified & \it Siloed & \it Unified \\ \hline\hline
        \textbf{TTFT (s)} & 14.5 & 12.9  & 34.9 & 34.5 & 1.0 & 1.0 & 1.0 & 1.0\\\hline
        \textbf{E2E (s)} & 55.3 & 53.3 & 98.3 & 99.1 & 10.6 & 10.5 & 19.2 & 18.9 \\ \hline
    \end{tabular}
\end{table}

This consolidation is reflected in the higher memory utilization for the Unified heuristic (\autoref{fig:motivation-utilization}), while not sacrificing the SLA latency for IW, e.g., with the change in $P95$ TTFT staying within $12\%$ for Bloom and almost identical for Llama (\autoref{table:silo_vs_reactive}) 
Thus, both process the full trace within the SLA,
but Unified uses fewer resources and also donates $52$ instance-hours to spot, compared to Siloed.
The memory utilization of Llama 2 is generally less than Bloom, indicating over allocation for Llama models. The Unified pool can take better advantage of this by re-dedicating GPUs from Llama to Bloom (inter-model scaling) and adapt better to the complementary demands compared to the Siloed approach, which does not allow allocation of VMs across models.
Hence, using a unified pool of instances for IW and NIW can improve resource utilization and cost efficiency, opening new optimization avenues for flexible NIW processing.
However, reactive scaling used by itself can affect the SLA of IW due to under-allocation, or raise the costs due to over-allocation (\autoref{fig:ideal_scaling}). It is is also sensitive to fine-grained temporal variations (\autoref{fig:wl:all1h}). This motivates the need for \textit{predictive scaling} that leverages TPS predictions for the workloads 
coupled with the Unified resource pool (\autoref{fig:arch_predict}), complemented by enhanced load balancing within a region.
Next, we formulate this as an optimization problem 
(\autoref{sec:optimization}) and then describe the architecture and heuristics of \sys (\autoref{sec:arch}).

%% file: SC_25/Optimization.tex
\section{Optimization Problem}
\label{sec:optimization}
We define an optimization problem for inference request routing and capacity allocation, to serve fast and slow workloads within the required SLA while maximizing utilization.

\para{Definition}~\textit{
Given a captive set of VMs of specific types in multiple regions, 
\begin{itemize}[noitemsep,topsep=0pt,parsep=0pt,partopsep=0pt]
    \item we need to continuously ensure that the right number of model instances of different model types are provisioned as endpoints at the regions, and 
    \item route the incoming workload across these endpoints,
\end{itemize}
such that
\begin{itemize}[noitemsep,topsep=0pt,parsep=0pt,partopsep=0pt]
    \item we maximize the utility of the workload requests completed within their SLA, and 
    \item maximize the capacity utilization of the VMs for the interactive workloads. 
\end{itemize}
}

We have two parts to solving this. 
\textit{First,} we need to \textbf{optimally provision instances} for model endpoints in different regions to handle this workload, within the available VM capacities. We also need to consider the overhead for (re)provisioning an LLM instance set onto VMs.
So, VM reprovisioning is only viable at a coarse granularity, e.g., every 15~mins, reclaiming spot instances can be fast, e.g., each 1~min.
\textit{Second,} we need to \textbf{route requests to the model instances} in different regions while meeting the SLA. These routing decisions can use real-time information on the load and responsiveness of the region endpoints.
We use tools like the SplitWise simulator~\cite{patel2024splitwise} to estimate the latency for serving requests at a certain request rate to ensure we meet the SLA, and ARIMA for load predictions~\cite{shumway2017arima}.
Spare model instances can be released to spot.
Next, we define these as an optimization problem. \autoref{tab:ILP_var} has the notations.

\begin{table}[t]
\vspace{-0.1in}
\small
\setlength{\tabcolsep}{2pt} 
    \centering
   \caption{Variables used in optimization problem}\vspace{-0.1in}
    \label{tab:ILP_var}
    \begin{tabular}{ccp{3.7cm}}
    \hline
        \bf Symbol & \bf Type  & \bf Description \\
         \hline\hline
$l$ & $\texttt{int}$ &  Model types  \\
$r$ & $\texttt{int}$ & Number of regions\\
$g$ & \texttt{int} & GPU types \\
$n_{i, j, k}$ &$[\texttt{int}]_{l\times r\times g}$ & Instances of model $i$ at region $j$ running on GPU $k$\\ 
$\rho_{i, j}(w)$ & $[\texttt{int}]_{l\times r}$ & {TPS requested} for model $i$ from clients at region $j$ for future time window $w$ \\ 
\hline
    \end{tabular}
\quad
\begin{tabular}{ccp{3.7cm}}
    \hline
        \bf Symbol & \bf Type  & \bf Description \\
         \hline\hline
         $\theta_{i\times k}$ & $ [\texttt{float}]_{l\times g}$ &  TPS provided by model $i$ on GPU $k$ \\
         $\alpha_{k}$ & $[\texttt{float}]_{g}$ & Cost of acquiring VM with GPU $k$ \\ 
         $\sigma_{i\times k}$ & $[\texttt{float}]_{l\times g}$ &  Cost of starting an instance of model $i$ on GPU $k$ \\
    \hline
        $\delta_{i, j, k}$ & $[\texttt{int}]_{l\times r\times g}$ & ILP output: optimal number of changes in VM allocation\\
    \hline

\end{tabular}
\vspace{-0.1in}
\end{table}

\para{Constraints} 
Say, the current number of VMs with GPU $k$ assigned to a model $i$ at region $j$ at a given time is $n_{i,j,k}$. Let $\delta_{i, j, k}$ be the optimal number of changes to be made to this VM count to service all IW requests in the next hour.
When \textit{servicing IW within a region},
say the forecasted TPS from clients
for model $i$ at region $j$ during each time window $w$ over the next decision making window of, say, 1~hour, is $\rho_{i,j}(w)$.
Say each model type in a region must serve at least $0 < \epsilon \leq 1$ fraction of its peak future request load 
in real-time. Excess load $(1-\epsilon)$ can be rerouted to other regions to reduce the number of model-instance changes needed. This is given by:
$\sum_{k}(n_{i,j, k} + \delta_{i, j, k}) \times \theta_{i, k} \geq \max_{w}{\ \epsilon \times \rho_{i,j}(w)} \ \forall i, j$.
When \textit{servicing IW across all regions,} 
we must ensure that all requests for a model $i$ received from all regions $j$ can cumulatively be processed using its model instances across all region, with rerouting, i.e.,
$\sum_{j}\sum_{k} (n_{i, j, k} + \delta_{i, j, k}) \times \theta_{i}\geq\max_{w}\sum_{j}\rho_{i,j}(w) \ \forall i$.
Also, we should never deallocate more models than are allocated, $\delta_{i, j, k} \geq -n_{i, j, k}$.

\para{Objective} Subject to the above constraints, we minimize the wasted resource overheads when provisioning VMs and instances required  for IW workloads while meeting their SLAs. We incur two overheads when provisioning a new model instance: (i) VM start up cost to instantiate a new VM ($\gamma$), and (ii) the deployment cost ($\mu$) when loading the model and its weights on a VM,
where $\gamma = \sum_{k}\Big(\alpha_{k} \times \sum_{i, j} \delta_{i, j, k} \Big)$ and $\mu = \sum_{k}\sum_{i}\sum_{j}\big(\sigma_{i, k} \times \max(0, \delta_{i, j, k}) \big)$.
These new VM are unusable during this period,
making 
spot instances more attractive. Given these, our objective is: $\boldsymbol{\arg \min (\gamma + \mu)}$.
Our formulation accounts for multiple models and regions, while allowing 
region re-routing or using GPUs with different throughputs for cost optimization. We assign values to $\theta_{i, k}$ by benchmarking model $i$ on GPU $k$, $\alpha_k$ using publicly available GPU VM costs and $\sigma_{i, k}$ as the product of $alpha_k$ with the average start up time of the model $i$ on GPU $k$.

\para{Finding Optimal Solutions} Since all the decision variables such as \# of changes in VM count are integers, we can solve this using ILP. While the runtime of an ILP solver can grow exponentially with the variables, we expect these parameters to be tractable.
E.g., the average solver runtime in our experiments with $l=4$, $r=3$ and $g=1$ was 1.41s. Increasing this with $l$ and $r$ to $20$ and $g$ to $5$ raises the solver time to 33s. These are acceptable for hourly decisions.

%% file: SC_25/architecture.tex
\section{Architecture of \sys}
\label{sec:arch}
Our \sys LLM serving architecture (\autoref{fig:simulator})
provides APIs similar to other LLM serving platforms, including real-time streaming API for IW that returns outputs once each token is
generated~\cite{streamAPI,aoai-stream} while using a Batch APIs~\cite{BatchAPI,aoai-batch} for NIW requests, which takes requests and returns responses asynchronously. The global router receives a request and routes it to the relevant region hosting the LLM instances which can service the request (\S~\ref{sec:instancerouter}). Others orchestrate the forecasting and optimization logic (\S~\ref{sec:forecast}), scaling logic (\S~\ref{sec:scaling} and NIW queue manager (\S~\ref{sec:queuemanager}). 

\subsection{Routing Logic}\label{sec:instancerouter}
The routing module of \sys (\autoref{fig:simulator}) decides routing of requests to regions by the \textit{Region Router}, routing to endpoints within a region, and routing locally among the instances of an endpoint.

\para{Global Routing for IW Requests}
When IW requests are received at the Region Router, the region routing logic checks the effective memory utilization of all the available regions hosting the model and routes the request to the region with memory utilization less than a given threshold, e.g., 70\%,
based on production systems at \csp. 
If multiple regions match, we can specify an order of preference, e.g., based on network proximity.
The \textit{effective memory utilization} is calculated as the ratio of the \textit{sum of the effective memory utilized} to the \textit{effective memory available} across all instances for a model in a region. The effective available memory for a model instance is obtained by subtracting model weights from the total VM memory. If none of the preferred regions has memory utilization less than the threshold, then the region with the least memory utilization among the choices is selected.

\para{Global Routing for NIW Requests}
NIW requests are sent by the Region Router to a \textit{Queue manager} that holds these requests (\autoref{fig:simulator}). Each model endpoint in a region periodically send a signal to the Queue Manager when their effective utilization falls below a specific threshold (60\%, in our experiments). Then, the NIW requests waiting in the queue for that region and model are incrementally removed by the Queue Manager and routed to the available endpoint (\autoref{sec:queuemanager}).

\para{Routing Logic at Region Endpoint} For both IW and NIW requests, the \textit{Model Router} routes requests arriving at a region to the least loaded deployment endpoint for that model, based on the effective memory utilized. Requests are sent to the instance with the minimum remaining tokens to process, based on the ``Join The Shortest Queue Logic''~\cite{gupta2007analysis}.

\para{Scheduler at the Model Instance} A local queue is maintained at the instance as well, and its scheduler batches inference requests in a first-come, first-served manner. IW requests arriving at an instance are assigned priority 0 and available immediately for inference. 
NIW requests may also arrive with a priority 0, as set by the Queue Manager when their deadline is approaching, and are considered on par with IW requests. Otherwise, NIW requests have a default priority 1, and are selected for inference only if there are no priority 0 requests ahead in the queue. 

\subsection{Queue Manager for NIW} \label{sec:queuemanager}
The \textit{Queue Manager} (\autoref{fig:simulator}) asynchronously routes NIW requests to specific regions and endpoints. Upon receiving a capacity availability signal from a model endpoint, async feedback logic pulls queued requests for that model type and routes them to the signaling region. NIW requests have a default 24-hour deadline. Requests with age $<10$ hours receive priority 1, while those $>10$ hours receive priority 0, similar to IW requests. If the signal indicates regional endpoint effective memory is less than 60\%, one request is sent; if less than 50\%, two requests are sent. These thresholds are tunable hyper-parameters. NIW requests with higher token counts can interfere with IW request execution, which is handled by chunking requests into fixed-size jobs. Therefore, we assume NIW request token counts are comparable to other workload tiers.

\subsection{Forecast Logic and Optimization Module}\label{sec:forecast}
For IW requests, the \textit{Load Predictor} (\autoref{fig:simulator}) forecasts the input TPS per region and per model type using the popular ARIMA time-series forecasting model~\cite{shumway2017arima}. As discussed in \S~\ref{fig:wl}, IW workloads are predictable and we find ARIMA to be accurate enough to forecast the diurnal load for each model in a region. We take the maximum TPS expected in the next hour from the forecast to estimate the TPS capacity required by the model to serve the IW load. We add a buffer of $\beta$ to this forecasted TPS to handle transient bursts and to offer spare capacity for NIW requests. We set the buffer as $\beta=10\%$ of the NIW load received in the past hour. The output of the forecast is sent to the optimization module, which uses an ILP solver to solve the optimization problem (\autoref{sec:optimization}).
This returns $\delta_{i,j}$, the change in the number of instances assigned to model $i$ at region $j$, which is passed to the Scaling Logic. The forecast and optimization modules run each hour.

\para{Choice of Forecast Model:} We choose an ARIMA model for forecasting the input TPS. We also explore four alternate models, with ARIMA providing accurate predictions with a low training and prediction latency, crucial to the performance of our system. We discuss these in~\autoref{app:forecasting}.

\subsection{Scaling Logic} \label{sec:scaling}

The long-term scaling logic, \textit{Autoscaler} (\autoref{fig:simulator}), uses the hourly 
instance change recommendations, $\delta_{i,j}$, for a specific region and model type to \textit{scale out} the instances, if $\delta_{i,j}>0$; and \textit{scale in} instances if $\delta_{i,j}<0$. For scale out, we first reclaim spot instances of the identical model type which are faster to acquire, and if none are available, we reclaim spot instances from other model types which can be slower to provision. Similarly, for scaling down, we donate the instances to the spot instances of the same model type.
We have a few choices on when to initiate the scaling.
\label{sec:lt-scaling-when}

\para{Immediate (LT-I)}
One naive approach is to instantly scale to the instance count recommended by the scaler every hour, which we call Immediate (LT-I).
However, the recommendation is based on the peak load that will occur in the next 1~hour. So scaling out immediately can cause transient over-provisioning, well before the peak load actually arrives.
So, we offer two additional deferred strategies to pace the rate at which the deployment state catches up with the forecasted load. 
These can improve the utilization of VMs and utility of serving requests, making spare capacity available to other models that need it in the same region.

\para{Deferred -- Instance Utilization (LT-U)}
We scale out only when the effective memory utilization actually starts increasing and goes over a threshold of 70\% used in our experiments. We keep increasing the instance count as long as this threshold is breached and until we achieve the instances count suggested by the optimization model. Similarly, when downscaling, we do so when the utilization goes below a 30\% threshold, again until we have reduced to the suggested number of instances. The region endpoint reports the effective memory usage when it receives a new request.

\para{Deferred -- Instance utilization and ARIMA gap (LT-UA)}
Our optimization model can make erroneous recommendations if the ARIMA predictions are inaccurate, which can happen with bursty requests. As for LT-U, in LT-UA we defer the scale out or in based on the memory usage thresholds actually being breached in either direction. 
However, we do not strictly stop the scale out or in if the instance count reaches the target count. Instead, during the last $20$~mins of the hour, if we have reached the scale out instance count and the observed TPS load is $\geq 5\times$ of what ARIMA predicted, then we continue to scale up the instance count (ARIMA highly underestimated). On the other hand, if the TPS load received is $\leq 0.5\times$ of what the ARIMA prediction, we continue scaling down (ARIMA highly overestimated). So, we switch from a memory-based strategy to a traffic-based strategy based on observed demand.

\subsection{Scheduling Logic}
The \textit{Request Scheduler} (\autoref{fig:simulator}) differentiates between fast and normal IW tiers to optimize the SLA. We define $d_r$ as the remaining time until the TTFT deadline for request $r$, where $d_r < 0$ indicates an expired deadline. The scheduler iterates through requests in the waiting queue, adding as many as possible to the current batch based on available GPU memory. The iteration order determines the request priority.
We evaluate four scheduling policies for request ordering:

\textit{First Come First Serve (FCFS)} orders requests by arrival timestamp, serving earlier arrivals first regardless of SLA tier or deadline urgency. This policy serves as our baseline.
\textit{Earliest Deadline First (EDF)} sorts requests by their $d_r$ values in ascending order. Requests with expired deadlines ($d_r<0$) are prioritized to prevent starvation. This naturally places IW-F requests ahead of IW-N requests arriving simultaneously, given IW-F's stricter TTFT.
\textit{Priority First (PF)} processes all IW-F requests in FCFS order before considering any IW-N requests. This policy maximizes IW-F user experience by ensuring absolute priority over IW-N requests.
Lastly, \textit{Deadline and Priority Aware (DPA)} provides a configurable policy that balances service quality across both tiers. The algorithm partitions requests into four deadline-based categories: (i) \textit{severely expired} requests with $d_r < -\tau_n$,  (ii) \textit{recently expired} requests with $-\tau_n\leq d_r < 0$, (iii) \textit{urgent requests} approaching deadline with $0\leq d_r \leq \tau_p$ and (iv) \textit{non-urgent requests} with $d_r > \tau_p$. DPA scheduling priority follows this order: (1) severely expired requests to prevent starvation, (2) urgent IW-F requests, (3) urgent IW-N requests, (4) non-urgent IW-F requests, (5) non-urgent IW-N requests, and finally (6) recently expired requests. The threshold parameters $\tau_n$ and $\tau_p$ enable fine-tuning the balance between tier prioritization and fairness.
While this is illustrated for two IW tiers, a similar approach can be extended to additional such latency tiers as well, in future.


%% file: SC_25/evaluation.tex
\section{Evaluation}
\label{sec:results}

Our detailed evaluation of \sys tests its ability to utilize GPU resources while maintaining latency targets for IW and NIW (\S~\ref{sec:overallperf},~\ref{sec:gpu_cost_sla_violation}); Compares it with SOTA scaling strategies on cost advantages (\S~\ref{sec:chiron_comparison},~\ref{sec:scaling_cost});
Evaluates it responsiveness and SLA maintenance for load bursts and prediction errors (\S~\ref{sec:loadbursts});
    and scaling over a week-long period with diurnal patterns  (\S~\ref{sec:salability}).

\subsection{Implementation and Simulating Setup}
\label{sec:results:simulator}

\begin{figure*}[t]
\vspace{-0.1in}
    \centering
    \begin{minipage}[b]{0.58\textwidth}
    \vspace{-0.1in}
        \centering
        \includegraphics[width=0.5\columnwidth]{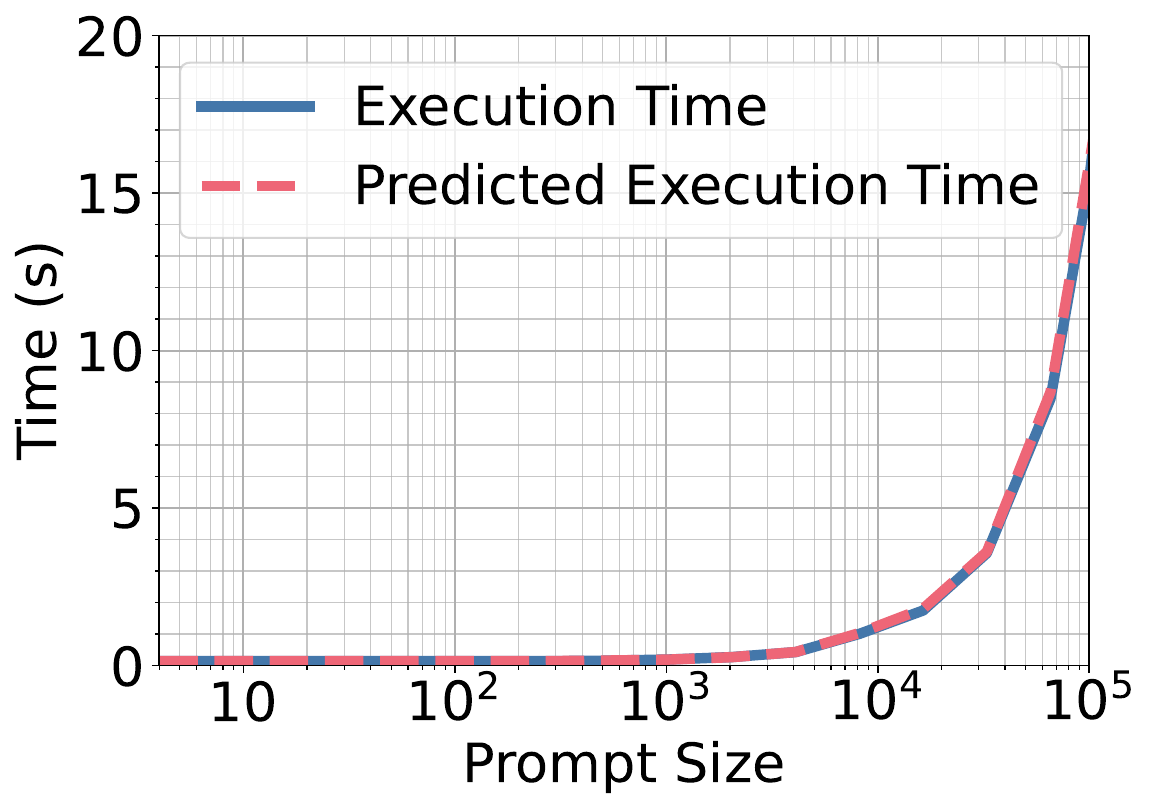}~
\includegraphics[width=0.49\columnwidth]{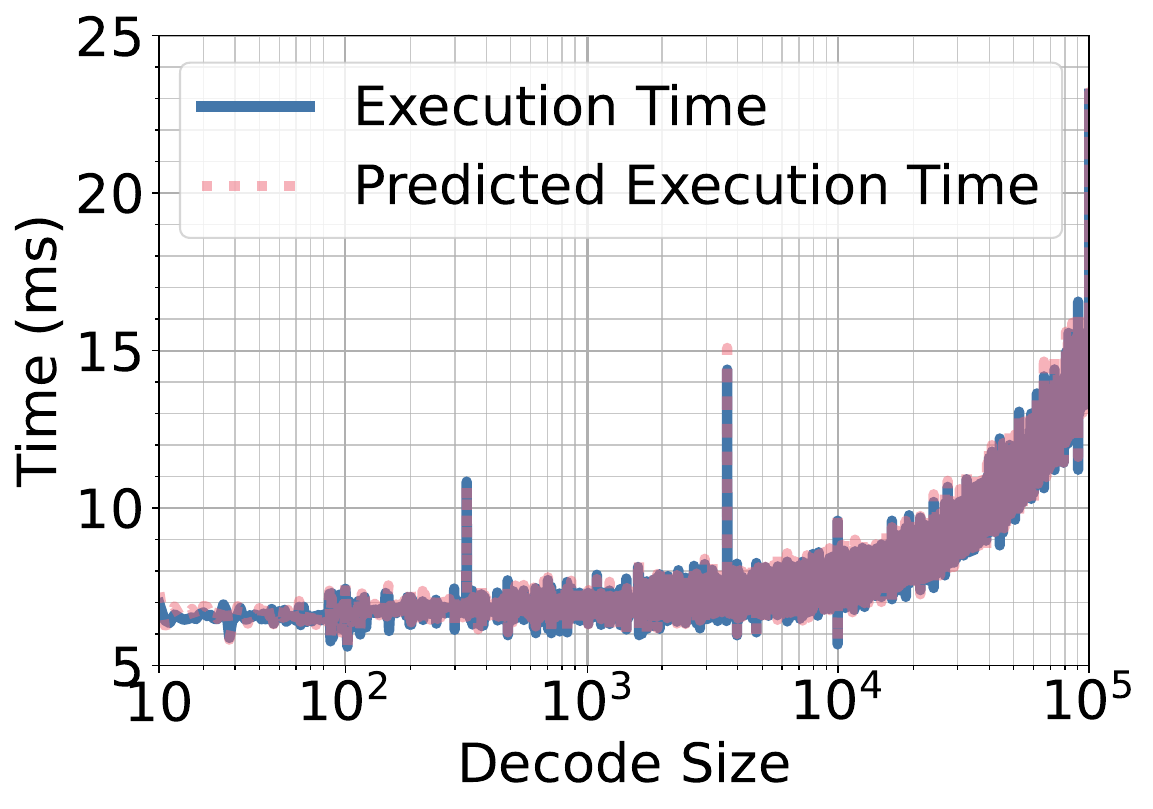}%
\vspace{-0.1in}
    \caption{Comparison of batch execution time predicted by Splitwise vs Real model instance for prompt phase (left) and decode phase (right). The $R^2$ value is 0.99 and 0.83 for prefill and decode phases respectively. We can infer from the plot that Llama 2 has a prompt TPS of 21000 for decode tokens.
    }
    \label{fig:splitwise_cs_actual}
    \end{minipage}\quad
    \begin{minipage}[b]{0.39\textwidth}
        \centering
        \includegraphics[width=1\columnwidth]{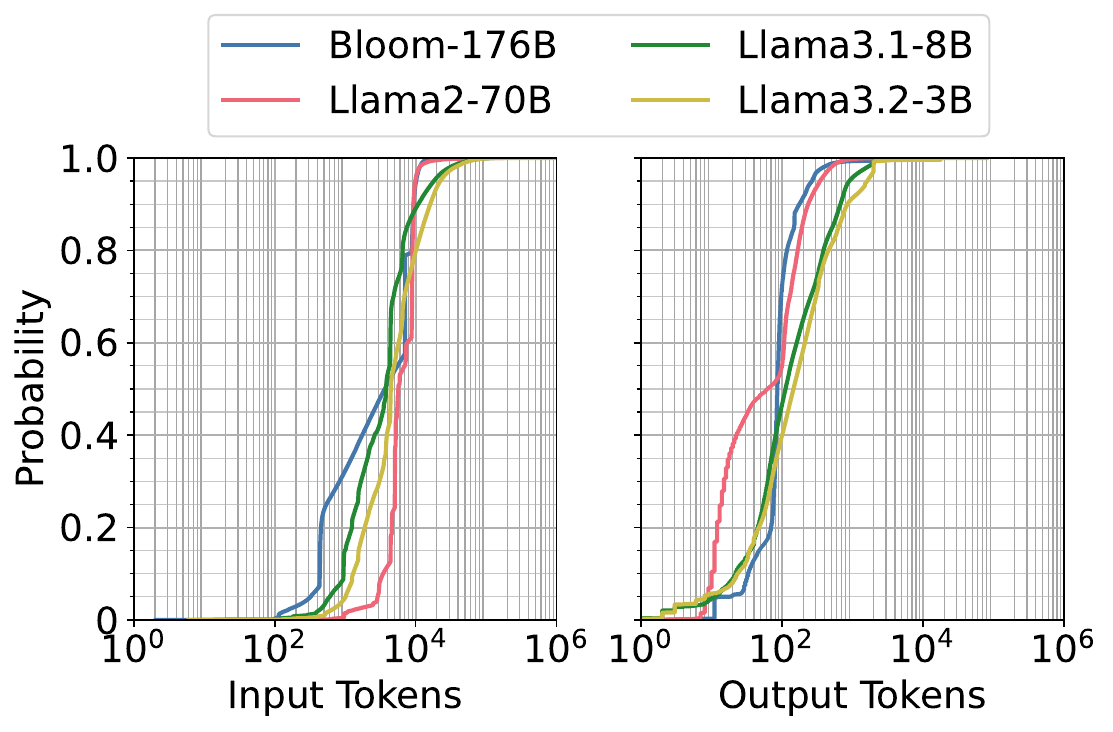}
        \vspace{-0.2in}
    \caption{CDF of Prompt, Output and Total Token Counts in log scale. Production traces for GPT models are mapped to open source LLMs used in evaluation.}
    \label{fig:cdf_tokens}
    \end{minipage}
    \vspace{-0.1in}
\end{figure*}

Experimenting with different scaling mechanisms, routing methods, and scheduling policies across multiple LLM types and GPU VMs can be costly in production. To enable scalable validation, we extend the SOTA LLM simulator \textit{Splitwise}~\cite{patel2024splitwise} to simulate our datacenter setup running multiple models on various hardware. Splitwise models components of a single LLM instance using Python-based discrete event simulation, including request queues. It uses a robust interpolation-based performance model from real inference traces to predict batch inferencing time per model, calculating user metrics like TTFT and E2E latencies. A single Splitwise instance equals one model instance deployment in a region. We verify simulator accuracy against real hardware deployments. Given the proprietary nature of O365's GPT models, we profile open-source models: Bloom-176B, Llama3.1-8B, Llama3.2-3B, and Llama2-70B on H100-80GB VMs with various input/output sizes~\cite{patel2024splitwise}. The simulator reports TTFT, TBT, E2E per request, and machine-level metrics using trained performance models. Estimates compared on an 80:20 train:test dataset split show MAPE of <3
<3\% (~\autoref{fig:splitwise_cs_actual}). Using this as an atomic model instance, we build our evaluation harness with multiple instances, a unified event queue, account routing, and model iterations. We further simulate multiple global regions to mimic \autoref{fig:simulator}. Our simulator is open-sourced for realistic evaluation of new routing and batching algorithms.

\para{Infrastructure Configuration and Hyper-parameters}
We select LLM infrastructure parameters in our simulations, such as utilization thresholds, to match production choices at O365, as discussed in the methods. The minimum instance count in a deployment is $2$, and the maximum is $3$. For around $90\%$ of regions, network latency is within $500ms$ with less than $2\%$ of cases having latency of $2.5s$. We select ARIMA hyper-parameters using AIC testing. ARIMA adds $\sim0.7s$ overhead for next-hour predictions and the ILP solver adds $\sim1.5s$ overhead for optimal instance allocation. \sys's intelligence is embedded within the controller module of the simulator (\autoref{fig:simulator}), which coordinates with different regions. Network overheads between regions affecting client request latency when routed to instances in different regions are captured using real latency distributions.

\para{GPUs and LLM Models} We evaluate workloads for the three US regions, and four standard LLM models, Bloom, Llama-2, Llama-3.1 and Llama-3.2, used by IW and NIW with their default weights. All the model types are assigned 20 instances per region at the start.
We assume homogeneous hardware and set the GPU cards needed by each model as identical. The TPS capacity for each LLM instance on each GPU type is shown in \autoref{fig:splitwise_cs_actual} 
We assume the redeployment time for an LLM model with weights available in the same region as 10 minutes for all the models regardless of their parameter size, while
the redeployment time of a model for which weights are absent in the local region is $\approx$2 hours. These are consistent with our observations from O365.
For spot instances being reclaimed, it takes a median of 1 minute and a maximum of 5 minutes. 
The simulator also handles unavailability of VMs being re-provisioned.

\para{Workload}
We use the July, 2025 trace by default in our experiments, but also confirm that the observed results are consistent with the Nov, 2024 traces.
\autoref{fig:cdf_tokens} provides a distribution of the input and output token counts for 4 internal models, which we map to the open source modes we evaluate. In general, the majority of requests have an input token count $>1k$, while most output token counts are $<1k$,  but vary with the model type.

\para{Baselines}
Unified Reactive Heuristic (\textit{Reactive}, \S~\ref{sec:empstudy}) baseline represents the current deployment at O365. Whenever a request arrives at the regional endpoint of a model type, the effective utilization is measured, and if the value is greater than 70\% ( < 30\%), the instance is scaled out (in),
with a cooldown period of 15s between scaling events. 
For SOTA baseline, we implement Chiron~\cite{chiron} in our simulator. We initialize the system with ten interactive, five mixed and five batch instances of each model type in every region and set $\Theta=0.6$ in its interactive autoscaling algorithm for ideal performance on our traces, as guided by their work. We keep the rest of the components like global routers and schedulers consistent between Chiron and \sys for a fair comparison.
We contrast the baselines with four strategies of \sys: LT-I, LT-U, and LT-UA.

\begin{figure}[t!]\vspace{-0.1in}
    \centering
\includegraphics[width=0.6\columnwidth]{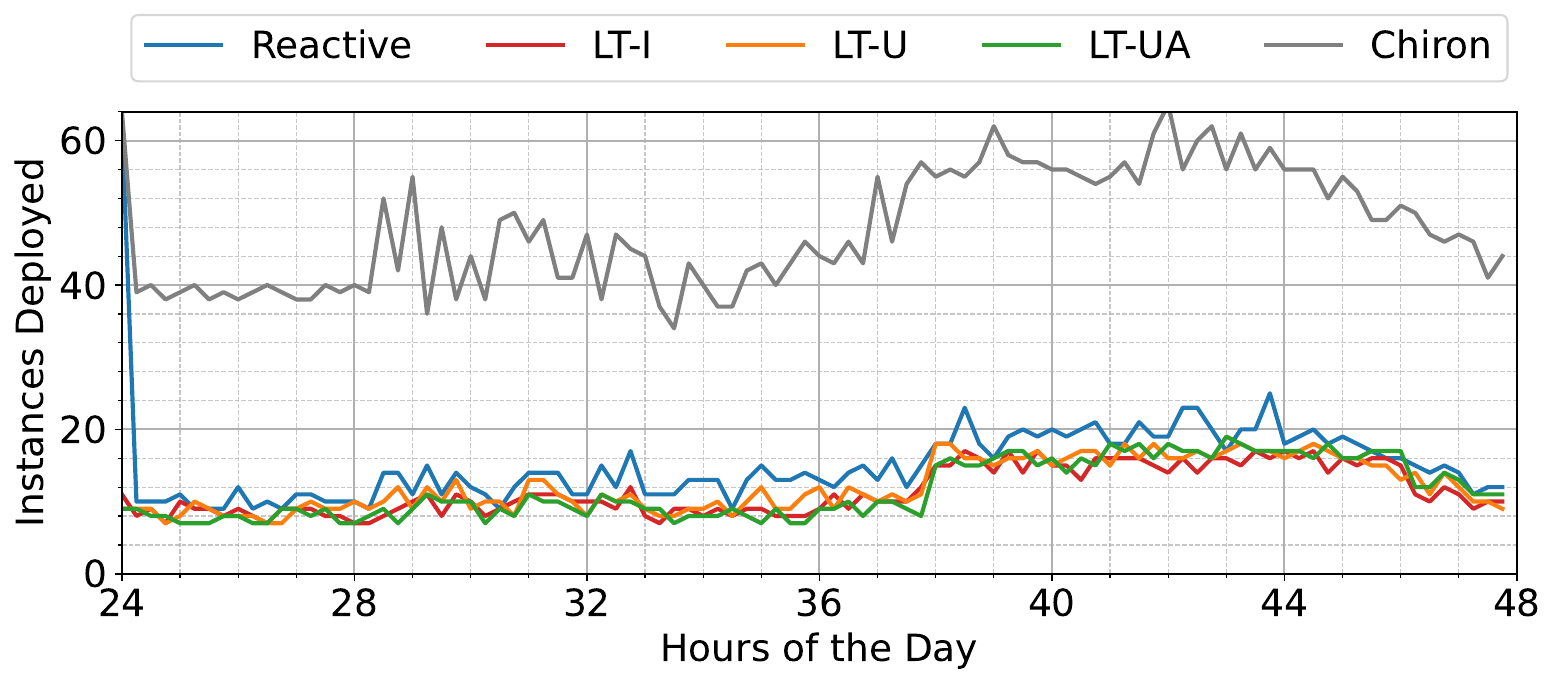}
\vspace{-0.1in}
    \caption{Aggregated sum of instances deployed across regions for Llama-2 on a peak traffic day.  
    Area under curve for Reactive, LT-I, LT-U, LT-UA and Chiron are 362.25, 274.5, 291, 277.5 and 1146, instance-hours. This translates to a \textit{savings of $\approx\$0.6M$} per week.
    }
    \label{fig:instance-hours-curve}
    \vspace{-0.1in}
\end{figure}

\begin{figure*}[ht]\vspace{-0.1in}
    \centering
    \begin{minipage}[b]{0.48\textwidth}
        \centering
        \subfloat[Model Instance Hours]{%
        \includegraphics[width=0.5\textwidth]{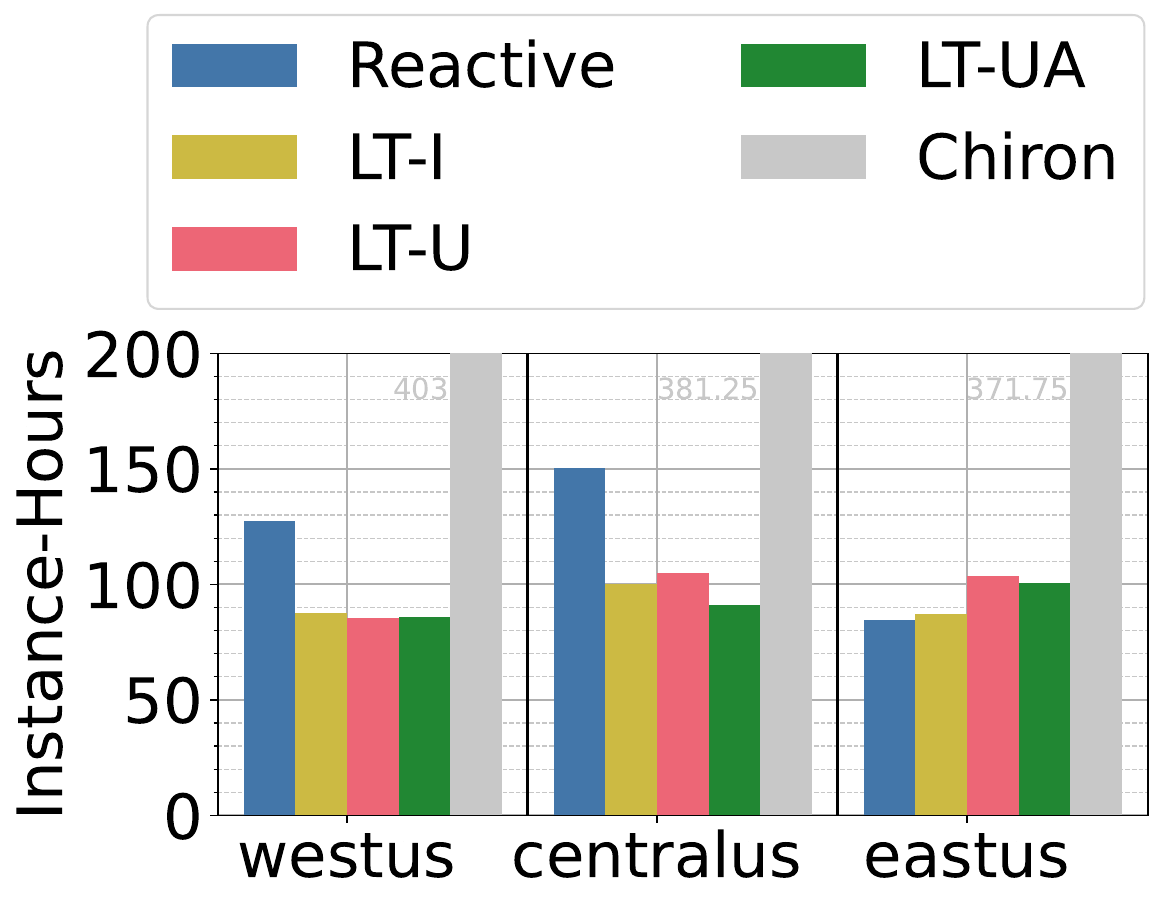}\label{fig:instance-hours-bar}}%
\hfill
    \subfloat[Memory Utilization]{
    \includegraphics[width=0.5\textwidth]{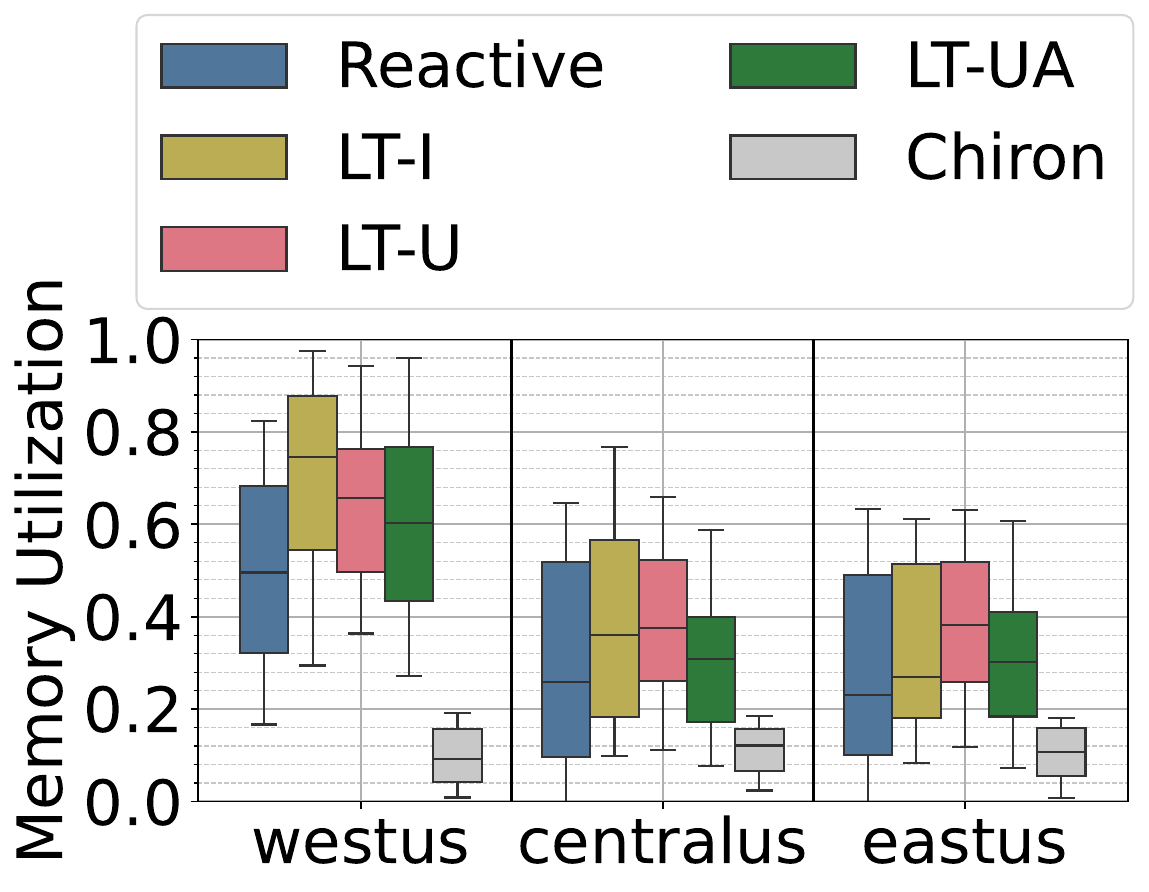}\label{fig:memory-util-box}%
    }
    \vspace{-0.1in}
    \caption{Llama-2 results for different strategies and regions.}
    \label{fig:motivation-mem-util-combined}
    \end{minipage}~
    \begin{minipage}[b]{0.48\textwidth}
        \centering
           \subfloat[75\%tile Latency Metrics]{
           \includegraphics[width=0.5\textwidth]{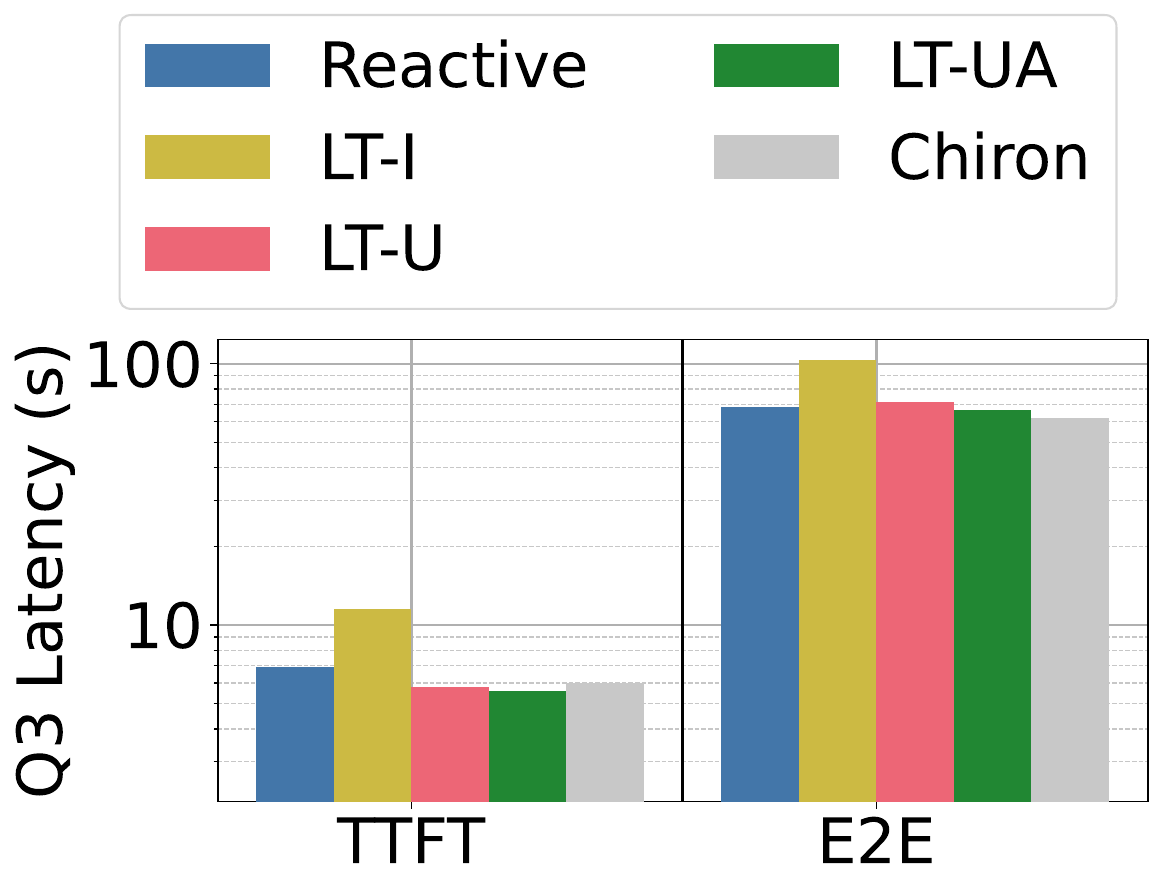}\label{fig:latency_metrics}}%
    \subfloat[GPU hours wasted on scaling]{
    \includegraphics[width=0.5\textwidth]{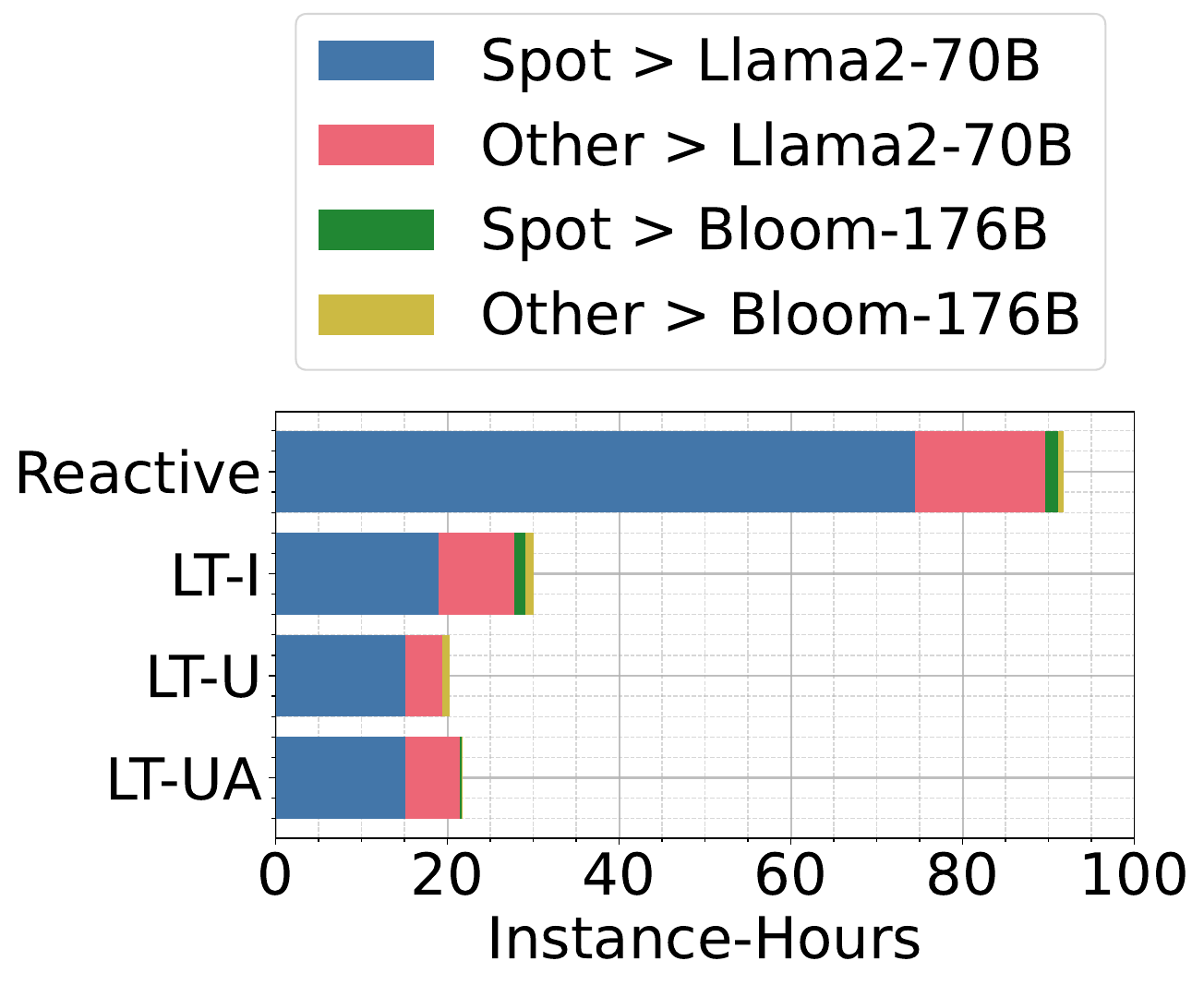}\label{fig:cost-of-scaling}}%
    \vspace{-0.1in}
    \caption{Llama-2 results on a peak traffic day. (b) shows the time to acquire spot instances.
    }
    \label{fig:motivation-mem-util-combined}
    \end{minipage}\vspace{-0.1in}
\end{figure*}

\begin{figure*}[t]
    \centering
    \begin{minipage}[b]{0.47\textwidth}
        \centering
       \subfloat[75\%tile TTFT]{
    \includegraphics[width=0.5\textwidth]{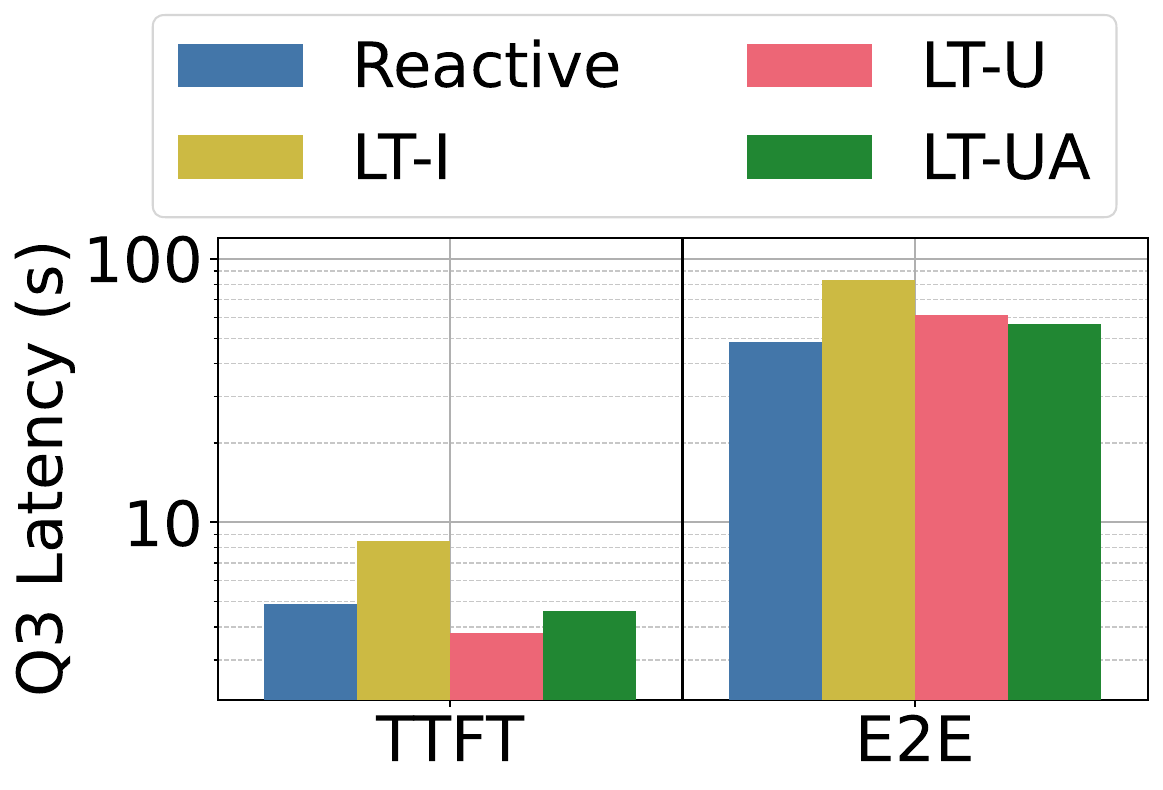}\label{fig:l4_latency}
    }%
    \subfloat[Instance Hours]{
    \includegraphics[width=0.5\textwidth]{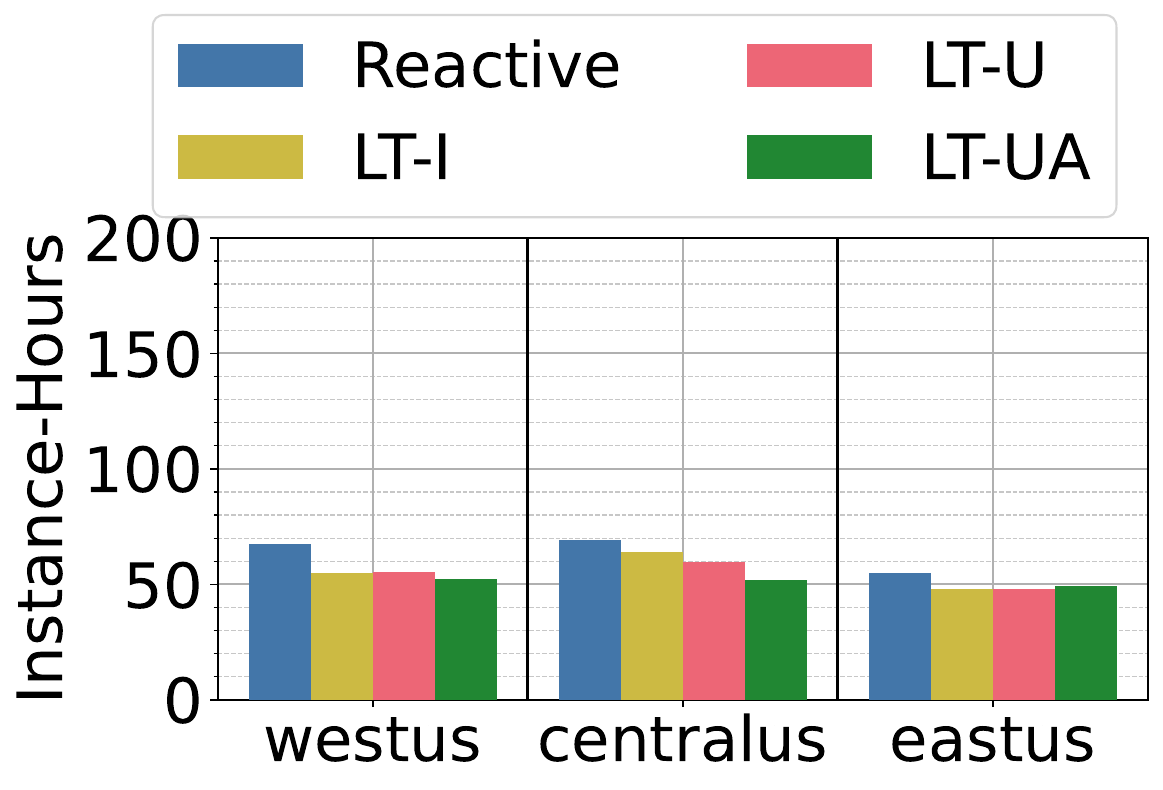}\label{fig:l4_instance_hours}
    }
    \vspace{-0.1in}
    \caption{Results after adding Llama-4 Scout as a fifth model to our experiments.}
    \end{minipage}~
    \begin{minipage}[b]{0.47\textwidth}
        \centering
            \subfloat[75\%tile TTFT]{%
    \includegraphics[width=0.5\textwidth]{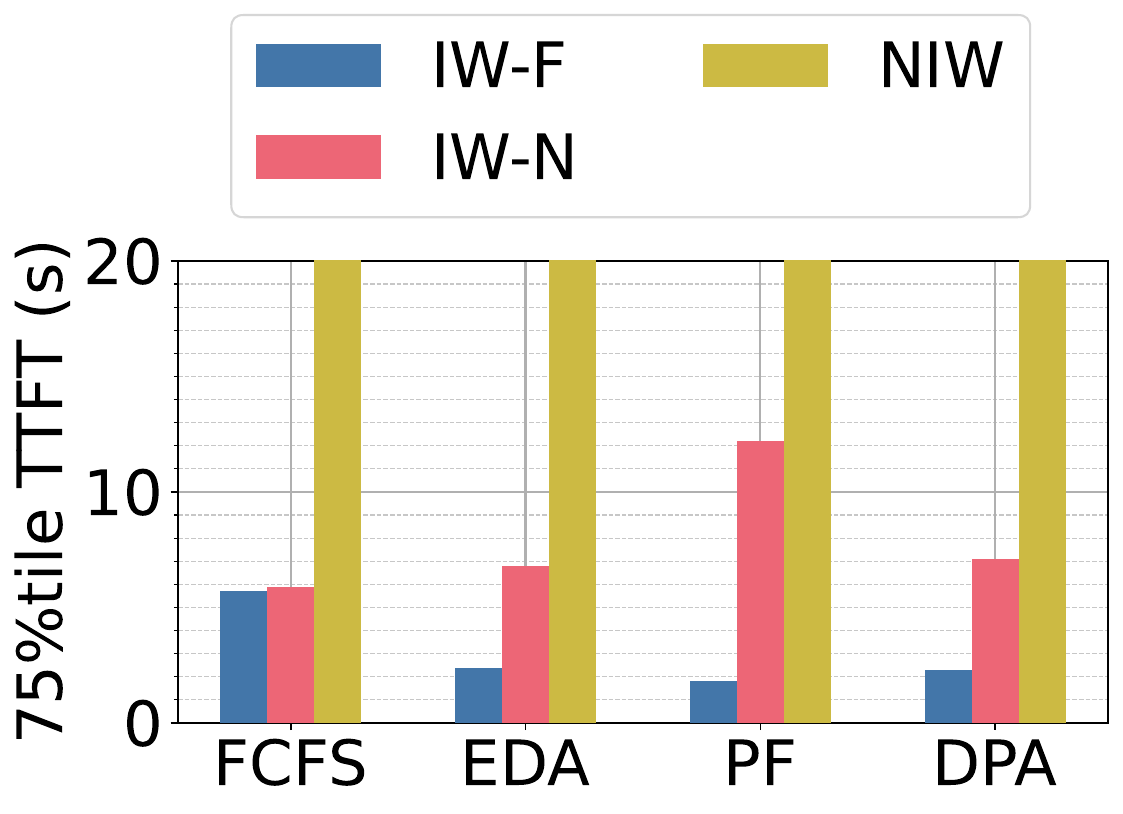}\label{fig:nw_ttft}%
    }%
    \subfloat[SLA Violations]{\includegraphics[width=0.5\textwidth]{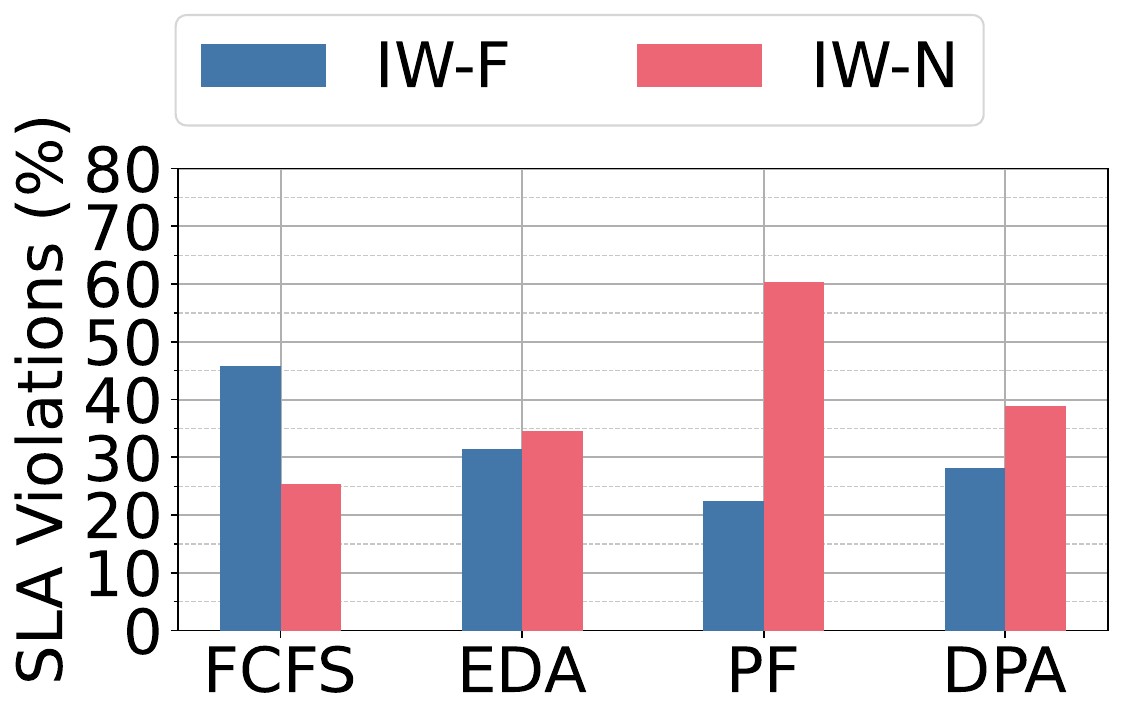}\label{fig:nw_sla_violations}}%
    \vspace{-0.1in}
    \caption {(a) Latency and (b) SLA violation for workloads as we chagne the scheduling policy.}
    \end{minipage}\vspace{-0.1in}
\end{figure*}

\subsection{Results}

\subsubsection{Effectiveness of Proactive Strategies in Reducing GPU-hours}\label{sec:overallperf}
We first evaluate the effectiveness of \sys for a single day trace. Figure~\ref{fig:instance-hours-curve} shows the trends of instance hour usage by hour, aggregated across all the three regions for Llama-2 70B model. Our forecast aware strategies consistently use less instances as compared to the reactive strategy, with LT-I, LT-U and LT-UA using 24.21\%, 19.65\% and 23.38\% less instance hours respectively. This is intuitive as LT approaches do not react to momentary bursts in traffic and scale based on the forecasts. These results are also evident in \autoref{fig:instance-hours-bar}, which shows LT strategies are better for all regions. 
At the time of writing, the cost of an H100 cluster on Azure is \$98.32/hour. Saving 85 instance hours a day for a single model in a single region would translate to roughly \$0.6M ( $\$98.32\times85\times3\text{ models}\times4\text{ regions }\times7\text{days}$) per week in our setting, or about \$2.5M per month!

\vspace{-0.07in}
\subsubsection{GPU Cost Reduction without SLO Violations}\label{sec:gpu_cost_sla_violation}
LT-I utilizes less GPU hours but it slightly harms the TTFT and E2E latency of requests, as evident from Figure~\ref{fig:latency_metrics}. This is because while immediately scaling up does not significantly benefit requests when traffic is lower, immediate scale down slows down requests, potentially harming their SLOs. These issues are fixed in LT-U and LT-UA, where instances scale on demand. These strategies help us downscale only if we can do so while serving low latency requests, while ensuring we do not up (or down) scale too much.

\vspace{-0.07in}
\subsubsection{Comparison with Chiron}\label{sec:chiron_comparison} 
Chiron optimizes solely for SLA, but can lead to increased instance demand without clear tail latency improvements (Figures~\ref{fig:instance-hours-curve} and~\ref{fig:latency_metrics}). Chiron also exhibits lower hardware utilization compared to us and even our reactive scaling mechanisms. This stems from its reliance on offline profiles for infrastructure scaling rather than online memory usage metrics.

\vspace{-0.07in}
\subsubsection{Scaling Costs}\label{sec:scaling_cost}
Due to the large amount of time for LLM instances to initiate even after GPU VMs have been acquired, frequent scale up events can result in increase in GPU-hours without any benefits to request serving.
In Figure~\ref{fig:cost-of-scaling} \sys is able to reduce the GPU cycles wasted during scale up events by about 70\%. Due to fluctuations in traffic, reactive scaling generally scale down unnecessarily, wasting GPU cycles when redeployment is inevitably needed. Using the forecasts, \sys reduces the number of times we upscale, resulting in better hardware usage.

\vspace{-0.07in}
\subsubsection{Scalability Test} 
We evaluate the generalizability of our \sys by incorporating a fifth 
Llama 4 Scout model into our experiment, with 109B parameters but utilizing a Mixture-of-Experts (MoE) architecture.
MoE's efficiency gains are reflected across our key performance metrics. Despite Scout's larger parameter count, we see substantial improvement in latency (\autoref{fig:l4_latency}) while maintaining high memory utilization. 
\sys's benefits
persist even after an increase in the baseline memory utilization of the Reactive approach. The higher throughput of Scout also results in a fewer instance hours for the model (\autoref{fig:l4_instance_hours}).
Hence, \sys scales effectively for diverse model architectures, adapting to unique efficiency characteristics of MoE models while maintaining the benefits for traditional dense models.

\vspace{-0.07in}
\subsubsection{Performance of Multi Tier Workloads}
Examining IW-F and IW-N, the default setting fails to distinguish between their different SLAs. Treating both the workloads similarly gives them similar Q3 TTFT ($\sim 5.6s$, \autoref{fig:nw_ttft}) but results in much higher SLA violations for IW-F ($\sim 45\%$) than IW-N ($\sim 25\%$, \autoref{fig:nw_sla_violations}). We test three different scheduling approaches of \sys to handle the SLA tiers. EDF scheduling balances the violations more evenly (31\% for IW-F and 34\% for IW-N) by reducing the Q3 TTFT for IW-F to 2.4s while increasing IW-N Q3 TTFT to 6.1s. PF scheduler achieves the lowest violations for IW-F (24\% with .9s response time) at the significant expense of IW-N (60\% violations and 12.1s TTFT). DPA provides a middle ground wwth 28\% and 38\% violations and 2.1s and 7.9s Q3 TTFT for IW-F and IW-N respectively. It can be further tuned to favor either workload types. All three schedulers can be extended to support additional SLA tiers. Cloud providers could implement SLA differentiation at the routing level as well by dedicating specific instances to high-priority workloads, though we leave this for future work.

\vspace{-0.07in}
\subsubsection{Burst Management using \sys}\label{sec:loadbursts}
\begin{figure*}
\vspace{-0.1in}
    \centering
    \subfloat[Burst Management]{
    \includegraphics[width=0.45\textwidth]{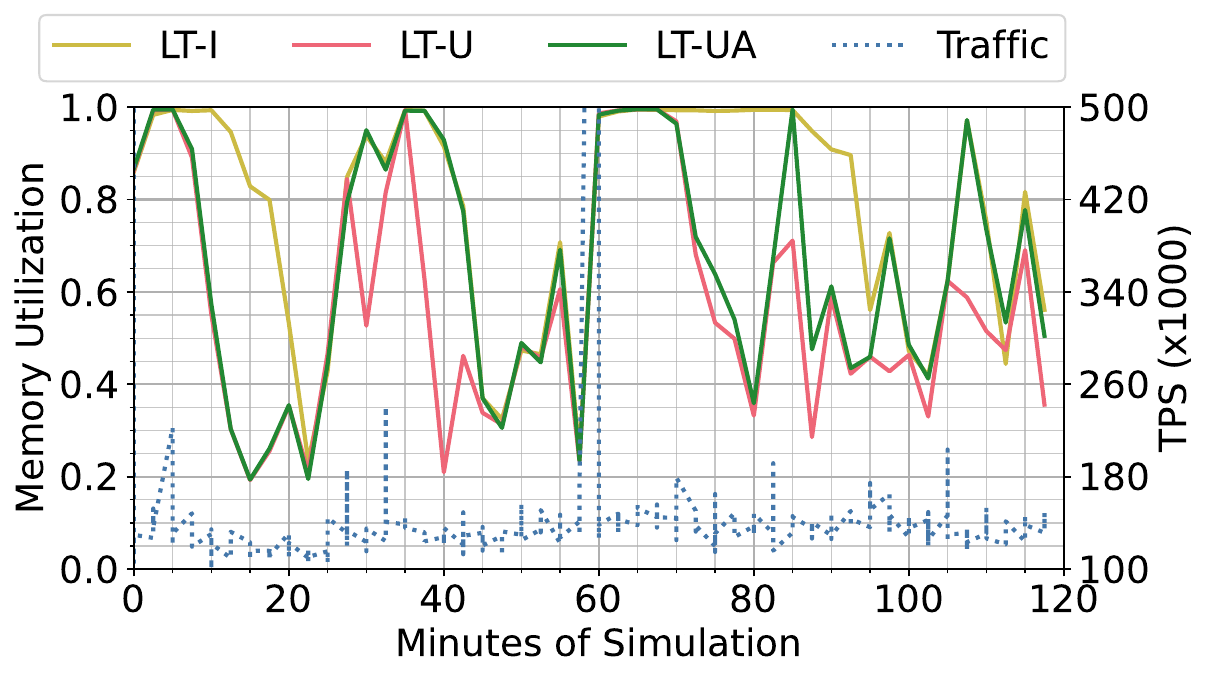}\label{fig:load-burst}
    }%
\hfill
    \subfloat[Weeklong 95\%tile Latency Metrics]{\includegraphics[width=0.4\textwidth]{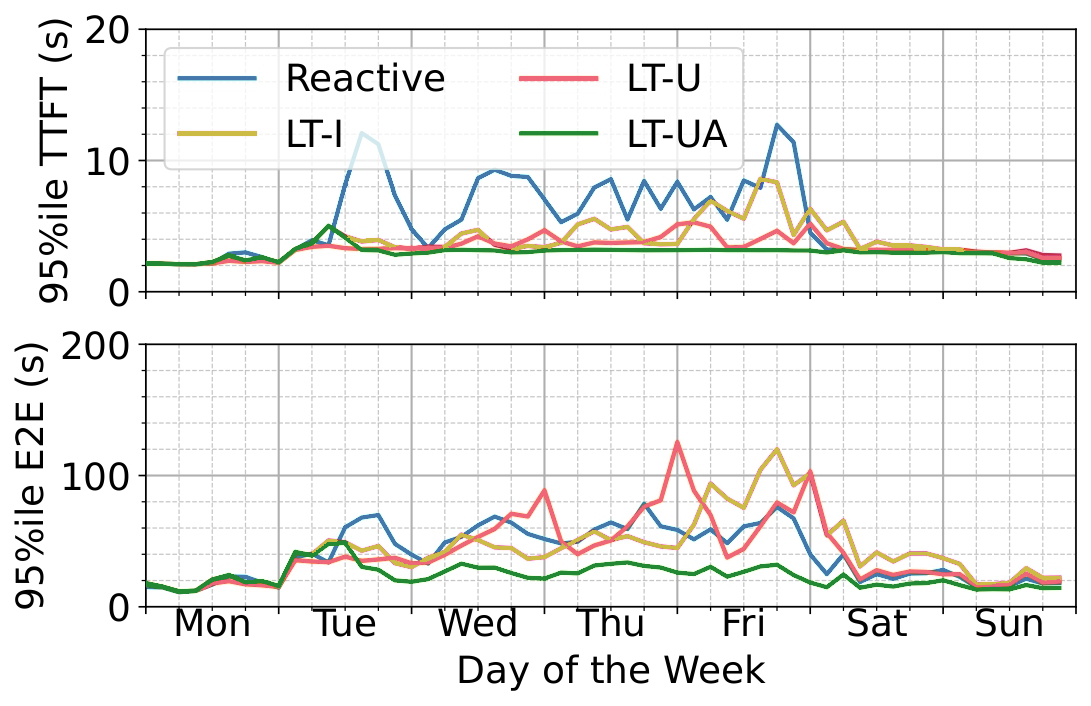}\label{fig:latency-curve}}
    \vspace{-0.1in}
    \caption{(a) Performance of LT-UA with synthetic request bursts.  (b) $P95$ latency binned by $3h$ for Llama2-70B.}
    \vspace{-0.1in}
\end{figure*}

We further evalaute the responsiveness to spikes by randomly increasing the incoming load to 8x to simulate sudden traffic bursts (blue curve, \autoref{fig:load-burst}). While LT-U and LT-I maintain their latency and memory usage for small bursts, they do not scale above the threshold set by the ILP and the ARIMA forecast even for large bursts. This is evident in the peak latencies during this time, where the green curve of LT-UA is able to reduce back it's memory usage faster than LT-I and LT-U. So, in such scenarios, LT-UA copes with the uncertainty much better. As discussed in \autoref{sec:lt-scaling-when}, we set the threshold to scale up at $5\times$ predicted traffic.

\para{Validation on Week Long Trace} \label{sec:salability}
\autoref{fig:latency-curve} displays the 95\%ile of TTFT and E2E latency over the course of one week. The insights gained from a one-day trace also apply in this case. The reactive strategy shows inferior performance, while other strategies achieve better performance metrics. LT-U and LT-UA behave similarly during weekdays, with a slight change in performance at the start of the weekend. This indicates the effect of the LT-UA strategy across longer time scales, which accounts for errors in ARIMA forecasts when the trend in TPS differs during the weekend. Overall, \sys scales well with longer traces and for different request rates.

\para{Validation for Nov, 2024 Trace}
We observe similar trends while evaluating using the earlier trace. Instance hours for Llama 2 on a peak traffic day are 302, 227, 248 and 233 for Reactive, LT-I, LT-U and LT-UA, with a 25\% reduction in instance hours seen for us. This validates the generalizability of our methods for production traces across time.

\para{Ablation Studies}
We test the robustness of our system in scenarios with varying load distribution and with different hardwares serving the LLMs as well. Our method generalizes well over these scenarios and we discuss them in~\autoref{app:ablation}.

%% file: SC_25/RelatedWork.tex
\section{Related Work}
\label{sec:related}
\para{Autoscaling for cloud computing}
There is a long history of forecast-based autoscaling for CPU VMs \cite{roy2011efficient,varshney2018autobot,hadary2020protean, rzadca2020autopilot}, illustrating its necessity in cloud services, challenges at cloud scale, and potential cost savings. For example, \cite{hadary2020protean} shows that even 1\% reduction in VM fragmentation yields \$100M annual savings, with higher expected savings for GPUs due to their greater cost \cite{patel2024splitwise}. Our work differs as we consider autoscaling LLMs on GPUs, which faces unique challenges: high latency sensitivity of interactive workloads, significant SLA variation across workloads, and high cost and latency of migrating and loading LLMs (hundreds of GB) on GPUs. Additionally, the unique nature of LLM computation with different prefill (compute bound) and decode (memory bound) phase characteristics must be considered when autoscaling.
\para{Efficient LLM routing and serving}
\cite{meta-llm-serving} discusses daily peaks, off-peaks, and unpredictable spikes in LLM inference workloads. Several works optimize latency or throughput at single model instances, including efficient attention computation \cite{dao2022flashattention, liuringattention}, batched inference \cite{orca, agrawal2024taming}, QoE optimization \cite{liu2024andes}, and fair service \cite{sheng2024fairness}. However, these consider only single LLM instances and do not account for multiple model types and GPUs.
\cite{jain2025performance} assumes identical LLM types and equal-priority workloads, focusing on load balancing across regional instances and addressing performance interference. \cite{fu2024serverlessllm} leverages additional storage and memory for faster LLM loading and live migration. While routing and scaling strategies exist for general workloads and LLMs, they often make simplifying assumptions.

\para{LLM autoscaling}
Since the advent of commercial LLM serving platforms \cite{ChatGPT, Copilot, Gemini}, research has explored autoscaling LLM resources to meet workload requirements. \cite{li2023alpaserve,melange,mei2024helix} explored optimal model placement on GPUs through various optimization formulations. However, these optimizations are static and do not consider dynamic workloads, rendering them ineffective for the high traffic fluctuations observed in commercial platforms (\autoref{fig:wl}).
Other works address specific serving challenges like pre-emption \cite{miao2024spotserve} and startup/migration delays \cite{fu2024serverlessllm}, but do not consider the diverse workload mix and varying model resource requirements that \sys does. Some works consider heterogeneous workloads but use reactive approaches: Llumnix\cite{llumnix} performs fine-grained context-switching for GPU memory defragmentation and load balancing, while ConServe\cite{qiao2024conserve} pre-empts NIW requests to prioritize interactive requests within single instances. Chiron \cite{chiron} performs backpressure-based autoscaling to optimize latency metrics for interactive requests across clusters with heterogeneous workloads.
These works focus on regional-level deployments, overlooking inter-region imbalances and load disparities across LLM types within regions. Using key workload insights, \sys proposes that scheduling and routing should consider the system holistically to better utilize available capacity. Finally, \cite{wu2023fast,nie2024aladdin} consider routing requests across model instances for latency minimization or throughput maximization. While request routing handles short-term traffic fluctuations, it must be combined with dynamic model scaling as in \sys to handle longer-term variations and avoid stale assignments.

\para{LLM workloads and simulators} While prior works such as Splitwise~\cite{patel2024splitwise}, Vidur~\cite{agrawal2024vidur}, and BurstGPT~\cite{wang2024burstgpt} have contributed datasets and simulators for LLM workloads, our simulator models the entire inference stack—from regions to deployments to models—within a heterogeneous environment. Unlike~\cite{patel2024splitwise}, which simulates only single model instances, and~\cite{agrawal2024vidur}, which lacks heterogeneous deployment and regional abstractions, our framework enables cloud-scale simulation across diverse hardware and model types. Datasets from prior works focus on narrow use cases like summarization~\cite{cohan2018discourse} or chat~\cite{wangopenchat,wang2024burstgpt,patel2024splitwise}, whereas our trace—sourced from Microsoft Copilot—captures enterprise LLM interactions across Microsoft365 apps like Word, Excel, PowerPoint, Teams, and Outlook, incorporating RAG-based inputs that diversify input and output distributions.
Additionally, datasets such as \cite{patel2024splitwise, agrawal2024vidur, wang2024burstgpt} are collected at regional levels and include only basic attributes like request time, input/output length, and LLM type. These limitations hinder capturing scaling challenges in realistic, global production environments. We extensively characterize workload tiers with different SLAs, their request-level characteristics, and the distribution of workloads and LLM types across multiple regions, using these insights to guide cost-effective solutions at scale.
Lastly, inference engines like vLLM~\cite{kwon2023efficient} and Sarathi-Serve~\cite{agrawal2024taming} can run LLMs on GPUs to serve workloads but are not designed for simulation-based capacity planning, which is central to our work.

%% file: SC_25/Conclusions.tex
\section{Conclusions}
\label{sec:conclude}
In this paper, we characterize LLM inferencing workloads from production traces of \csp, revealing insights on temporal behavior across latency tiers and regions. We use this to design \sys, a holistic system for serving LLM inference requests with diverse SLAs that maintains better GPU utilization, reduces resource fragmentation from silos, and increases utility by donating surplus instances to Spot instances. \sys achieves this through its holistic deployment stack for varying SLA requests, async feed module for NIW, and long-term aware proactive scaler logic that capitalizes on underutilized instances through inter-model redeployment. These benefits are confirmed through experiments with a realistic simulator and our traces against baseline and SOTA methods, potentially saving millions of USD monthly. 

As lessons learned, theoretical performance limits for LLM types differ from real-world achievements. Cost-effective solutions at scale require extensive performance benchmarking and deep production insights. Future work includes extending \sys to accommodate workloads with a continuum of SLAs and conducting studies on the proposed approach with deployments across heterogeneous hardware types.


%% file: sigmetrics/Appendix.tex
\appendix
\newpage
\begin{figure}[t]
    \centering
    \includegraphics[width=0.6\linewidth]{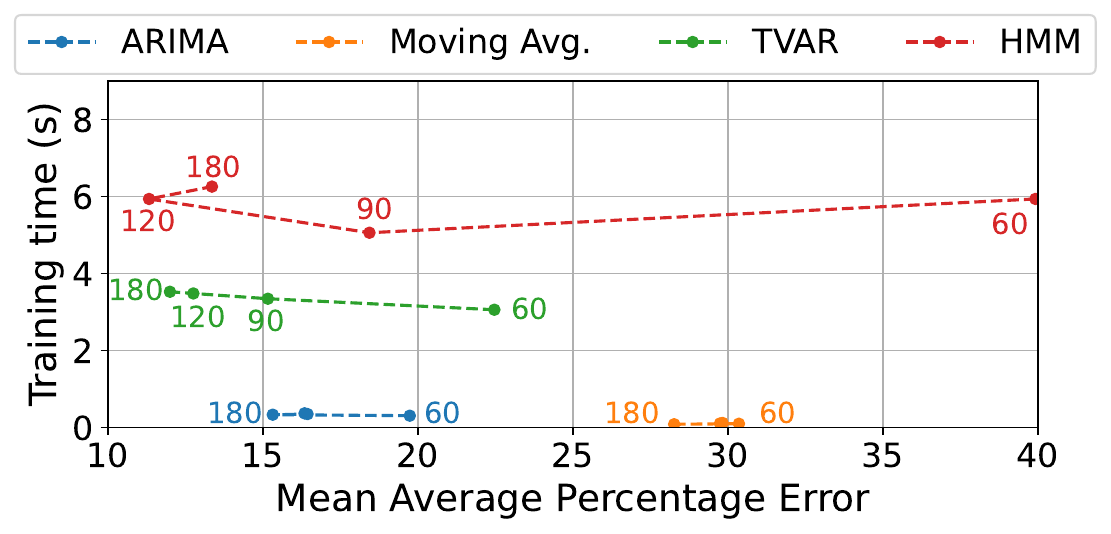}
    \caption{\textbf{Forecast Prediction Models}: Performance of four forecasting models in terms of their mean absolute percentage error and training time. We test these models by training them on four different training windows and study the impact on their training time and accuracy. The moving average method shows poor performance, while the Hidden Markov Method is quite expensive to train, making it an unsuitable choice despite its performance. Both, TVAR and ARIMA are suitable candidates for our design, being fast while being accurate enough. We choose ARIMA with a training window of 60 for our experiments due to its lower training latency.
    }
    \label{fig:forecast_prediction_models}
\end{figure}
\section{Ablation Studies}\label{app:ablation}
We conducted experiments with different hardware and different workload distributions to test the robustness of our methods. \sys (LT-I, LT-U, LT-UA) only requires the token processing profile to be extended to new model hardware pairs. For experiments on A100 clusters, our LT-UA approach uses 28.2\% fewer GPU-hours while maintaining tail latency metrics compared to Reactive scaling. This is mainly due to the longer model loading times on A100 clusters.
Next, we change the IW to NIW ratio in our traces. 
In the production traces from Nov. 2024 we see a 3:1 ratio of IW and NIW requests, but our methods are independent of this distribution. After changing this distribution to 9:1 and 1:1, we find that LT-UA requires 26.3\% and 22\% fewer GPU hours compared to Reactive scaling. These variations are expected, as LT-UA uses the NIW token count to determine the buffer size, which is higher than the original setting in the latter case and lower in the former.

\section{Forecast Prediction Models}\label{app:forecasting}
For incoming Tokens Per Minute (TPM) forecasting, we compare ARIMA with the following three additional models along with different window lengths. The training data horizon is 60 minutes.
\begin{enumerate}
    \item Moving Average: This is a simple moving average which predicts the next TPM to be the average of the previous window.
    \item Time Varying Autoregressive Model (TVAR): This is a popular autoregressive model, which trains on a user-defined window length and makes single value predictions at each time step.
    \item Hidden Markov Models (HMM): The date, time, and tokens received in the previous window are used as states of the Markov Model to predict the expected incoming TPM over a 60 step horizon.
\end{enumerate}
We compare these methods in their Mean Average Percentage Error in their prediction and latency to train the models.~\autoref{fig:forecast_prediction_models} shows that TVAR and HMM can achieve better prediction accuracy but take much longer training time than ARIMA, while the moving average is quite fast but inaccurate. TVAR with a training window of 90, 120, or 180 would have been a good choice for our system as well, however, we chose ARIMA due to relatively faster training time.

%% file: main.bib
@article{melange,
  title={M$\backslash$'elange: Cost Efficient Large Language Model Serving by Exploiting GPU Heterogeneity},
  author={Griggs, Tyler and Liu, Xiaoxuan and Yu, Jiaxiang and Kim, Doyoung and Chiang, Wei-Lin and Cheung, Alvin and Stoica, Ion},
  journal={arXiv preprint arXiv:2404.14527},
  year={2024}
}

@article{mei2024helix,
  title={Helix: Distributed Serving of Large Language Models via Max-Flow on Heterogeneous GPUs},
  author={Mei, Yixuan and Zhuang, Yonghao and Miao, Xupeng and Yang, Juncheng and Jia, Zhihao and Vinayak, Rashmi},
  journal={arXiv preprint arXiv:2406.01566},
  year={2024}
}

@inproceedings{fu2024serverlessllm,
  title={ServerlessLLM: Low-Latency Serverless Inference for Large Language Models},
  author={Fu, Yao and Xue, Leyang and Huang, Yeqi and Brabete, Andrei-Octavian and Ustiugov, Dmitrii and Patel, Yuvraj and Mai, Luo},
  booktitle={18th USENIX Symposium on Operating Systems Design and Implementation (OSDI 24)},
  pages={135--153},
  year={2024}
}

@inproceedings{liuringattention,
  title={RingAttention with Blockwise Transformers for Near-Infinite Context},
  author={Liu, Hao and Zaharia, Matei and Abbeel, Pieter},
  booktitle={The Twelfth International Conference on Learning Representations}
}

@inproceedings{sheng2024fairness,
  title={Fairness in serving large language models},
  author={Sheng, Ying and Cao, Shiyi and Li, Dacheng and Zhu, Banghua and Li, Zhuohan and Zhuo, Danyang and Gonzalez, Joseph E and Stoica, Ion},
  booktitle={18th USENIX Symposium on Operating Systems Design and Implementation (OSDI 24)},
  pages={965--988},
  year={2024}
}

@article{qiao2024conserve,
  title={ConServe: Harvesting GPUs for Low-Latency and High-Throughput Large Language Model Serving},
  author={Qiao, Yifan and Anzai, Shu and Yu, Shan and Ma, Haoran and Wang, Yang and Kim, Miryung and Xu, Harry},
  journal={arXiv preprint arXiv:2410.01228},
  year={2024}
}

@inproceedings{miao2024spotserve,
  title={Spotserve: Serving generative large language models on preemptible instances},
  author={Miao, Xupeng and Shi, Chunan and Duan, Jiangfei and Xi, Xiaoli and Lin, Dahua and Cui, Bin and Jia, Zhihao},
  booktitle={Proceedings of the 29th ACM International Conference on Architectural Support for Programming Languages and Operating Systems, Volume 2},
  pages={1112--1127},
  year={2024}
}

@inproceedings{kwon2023efficient,
  title={Efficient memory management for large language model serving with pagedattention},
  author={Kwon, Woosuk and Li, Zhuohan and Zhuang, Siyuan and Sheng, Ying and Zheng, Lianmin and Yu, Cody Hao and Gonzalez, Joseph and Zhang, Hao and Stoica, Ion},
  booktitle={Proceedings of the 29th Symposium on Operating Systems Principles},
  pages={611--626},
  year={2023}
}

@inproceedings{Llama,
  title={Llama 2 is here - Get it on Hugging Face [Online]},
  author={P. Schmid and O. Sansevieroa and P. Cuenca and L. Tunstall},
  booktitle={Available:
https://huggingface.co/blog/llama2}
}

@inproceedings{Bloom,
  title={Introducing The World’s Largest Open Multilingual Language Model: BLOOM [Online]},
  author={BigScience},
  booktitle={https://bigscience.huggingface.co/blog/bloom}
}

@misc{BatchAPI,
  title={Batch API},
  author={OpenAI},
  title={https://platform.openai.com/docs/guides/batch/batch-api},
year = {2024}
}

@misc{AzureBatchAPI,
  title={Azure OpenAI Batch API},
  author={Microsoft},
  title={https://learn.microsoft.com/en-us/azure/ai-foundry/openai/how-to/batch?tabs=global-batch%2Cstandard-input%2Cpython-secure&pivots=ai-foundry-portal},
year = {2025}
}

@misc{streamAPI,
  title={Streaming API},
  author={OpenAI},
  note={https://platform.openai.com/docs/api-reference/streaming},
year = {2024}
}

@article{gupta2007analysis,
  title={Analysis of join-the-shortest-queue routing for web server farms},
  author={Gupta, Varun and Balter, Mor Harchol and Sigman, Karl and Whitt, Ward},
  journal={Performance Evaluation},
  volume={64},
  number={9-12},
  pages={1062--1081},
  year={2007},
  publisher={Elsevier}
}

@misc{aoai-batch,
    title = {Run Azure OpenAI models in batch endpoints to compute embeddings},
    author={Microsoft},
year={2024},
    note = {\url{https://learn.microsoft.com/en-us/azure/machine-learning/how-to-use-batch-model-openai-embeddings}}
}

@misc{aoai-stream,
    title = {Online endpoint deployment for real-time inferencing},
    author={Microsoft},
year={2024},
    note = {\url{https://learn.microsoft.com/en-us/azure/machine-learning/concept-endpoints-online}}
}

@misc{ChatGPT,
  title = {ChatGPT},
  howpublished = {\url{http://chat.openai.com}}
}

@misc{Gemini,
  title = {Gemini},
  howpublished = {\url{http://gemini.google.com}}
}

@misc{Copilot,
  title = {Copilot},
  howpublished = {\url{http://copilot.microsoft.com}}
}

@article{varshney2018autobot,
  title={AutoBoT: Resilient and cost-effective scheduling of a bag of tasks on spot VMs},
  author={Varshney, Prateeksha and Simmhan, Yogesh},
  journal={IEEE Transactions on Parallel and Distributed Systems},
  volume={30},
  number={7},
  pages={1512--1527},
  year={2018},
  publisher={IEEE}
}

@inproceedings{roy2011efficient,
  title={Efficient autoscaling in the cloud using predictive models for workload forecasting},
  author={Roy, Nilabja and Dubey, Abhishek and Gokhale, Aniruddha},
  booktitle={2011 IEEE 4th International Conference on Cloud Computing},
  pages={500--507},
  year={2011},
  organization={IEEE}
}

@inproceedings{hadary2020protean,
  title={Protean:$\{$VM$\}$ allocation service at scale},
  author={Hadary, Ori and Marshall, Luke and Menache, Ishai and Pan, Abhisek and Greeff, Esaias E and Dion, David and Dorminey, Star and Joshi, Shailesh and Chen, Yang and Russinovich, Mark and others},
  booktitle={14th USENIX Symposium on Operating Systems Design and Implementation (OSDI 20)},
  pages={845--861},
  year={2020}
}

@inproceedings{rzadca2020autopilot,
  title={Autopilot: workload autoscaling at google},
  author={Rzadca, Krzysztof and Findeisen, Pawel and Swiderski, Jacek and Zych, Przemyslaw and Broniek, Przemyslaw and Kusmierek, Jarek and Nowak, Pawel and Strack, Beata and Witusowski, Piotr and Hand, Steven and others},
  booktitle={Proceedings of the Fifteenth European Conference on Computer Systems},
  pages={1--16},
  year={2020}
}

@article{shumway2017arima,
  title={ARIMA models},
  author={Shumway, Robert H and Stoffer, David S and Shumway, Robert H and Stoffer, David S},
  journal={Time series analysis and its applications: with R examples},
  pages={75--163},
  year={2017},
  publisher={Springer}
}

@inproceedings{li2023alpaserve,
  title={$\{$AlpaServe$\}$: Statistical multiplexing with model parallelism for deep learning serving},
  author={Li, Zhuohan and Zheng, Lianmin and Zhong, Yinmin and Liu, Vincent and Sheng, Ying and Jin, Xin and Huang, Yanping and Chen, Zhifeng and Zhang, Hao and Gonzalez, Joseph E and others},
  booktitle={17th USENIX Symposium on Operating Systems Design and Implementation (OSDI 23)},
  pages={663--679},
  year={2023}
}

@article{wu2023fast,
  title={Fast distributed inference serving for large language models},
  author={Wu, Bingyang and Zhong, Yinmin and Zhang, Zili and Liu, Shengyu and Liu, Fangyue and Sun, Yuanhang and Huang, Gang and Liu, Xuanzhe and Jin, Xin},
  journal={arXiv preprint arXiv:2305.05920},
  year={2023}
}

@article{nie2024aladdin,
  title={Aladdin: Joint Placement and Scaling for SLO-Aware LLM Serving},
  author={Nie, Chengyi and Fonseca, Rodrigo and Liu, Zhenhua},
  journal={arXiv preprint arXiv:2405.06856},
  year={2024}
}

@article{dao2022flashattention,
  title={Flashattention: Fast and memory-efficient exact attention with io-awareness},
  author={Dao, Tri and Fu, Dan and Ermon, Stefano and Rudra, Atri and R{\'e}, Christopher},
  journal={Advances in Neural Information Processing Systems},
  volume={35},
  pages={16344--16359},
  year={2022}
}

@inproceedings {orca,
author = {Gyeong-In Yu and Joo Seong Jeong and Geon-Woo Kim and Soojeong Kim and Byung-Gon Chun},
title = {Orca: A Distributed Serving System for {Transformer-Based} Generative Models},
booktitle = {16th USENIX Symposium on Operating Systems Design and Implementation (OSDI 22)},
year = {2022},
isbn = {978-1-939133-28-1},
address = {Carlsbad, CA},
pages = {521--538},
url = {https://www.usenix.org/conference/osdi22/presentation/yu},
publisher = {USENIX Association},
month = jul
}

@inproceedings{agrawal2024taming,
  title={Taming $\{$Throughput-Latency$\}$ Tradeoff in $\{$LLM$\}$ Inference with $\{$Sarathi-Serve$\}$},
  author={Agrawal, Amey and Kedia, Nitin and Panwar, Ashish and Mohan, Jayashree and Kwatra, Nipun and Gulavani, Bhargav and Tumanov, Alexey and Ramjee, Ramachandran},
  booktitle={18th USENIX Symposium on Operating Systems Design and Implementation (OSDI 24)},
  pages={117--134},
  year={2024}
}

@inproceedings{patel2024splitwise,
  title={Splitwise: Efficient generative llm inference using phase splitting},
  author={Patel, Pratyush and Choukse, Esha and Zhang, Chaojie and Shah, Aashaka and Goiri, {\'I}{\~n}igo and Maleki, Saeed and Bianchini, Ricardo},
  booktitle={2024 ACM/IEEE 51st Annual International Symposium on Computer Architecture (ISCA)},
  pages={118--132},
  year={2024},
  organization={IEEE}
}

@article{liu2024andes,
  title={Andes: Defining and Enhancing Quality-of-Experience in LLM-Based Text Streaming Services},
  author={Liu, Jiachen and Wu, Zhiyu and Chung, Jae-Won and Lai, Fan and Lee, Myungjin and Chowdhury, Mosharaf},
  journal={arXiv preprint arXiv:2404.16283},
  year={2024}
}

@misc{chiron,
      title={Hierarchical Autoscaling for Large Language Model Serving with Chiron}, 
      author={Archit Patke and Dhemath Reddy and Saurabh Jha and Chandra Narayanaswami and Zbigniew Kalbarczyk and Ravishankar Iyer},
      year={2025},
      eprint={2501.08090},
      archivePrefix={arXiv},
      primaryClass={cs.DC},
      url={https://arxiv.org/abs/2501.08090}, 
}

@inproceedings{llumnix,
author = {Sun, Biao and Huang, Ziming and Zhao, Hanyu and Xiao, Wencong and Zhang, Xinyi and Li, Yong and Lin, Wei},
title = {Llumnix: dynamic scheduling for large language model serving},
year = {2024},
isbn = {978-1-939133-40-3},
publisher = {USENIX Association},
address = {USA},
booktitle = {Proceedings of the 18th USENIX Conference on Operating Systems Design and Implementation},
articleno = {10},
numpages = {19},
location = {Santa Clara, CA, USA},
series = {OSDI'24}
}

@misc{cui2025curieevaluatingllmsmultitask,
      title={CURIE: Evaluating LLMs On Multitask Scientific Long Context Understanding and Reasoning}, 
      author={Hao Cui and Zahra Shamsi and Gowoon Cheon and Xuejian Ma and Shutong Li and Maria Tikhanovskaya and Peter Norgaard and Nayantara Mudur and Martyna Plomecka and Paul Raccuglia and Yasaman Bahri and Victor V. Albert and Pranesh Srinivasan and Haining Pan and Philippe Faist and Brian Rohr and Michael J. Statt and Dan Morris and Drew Purves and Elise Kleeman and Ruth Alcantara and Matthew Abraham and Muqthar Mohammad and Ean Phing VanLee and Chenfei Jiang and Elizabeth Dorfman and Eun-Ah Kim and Michael P Brenner and Viren Jain and Sameera Ponda and Subhashini Venugopalan},
      year={2025},
      eprint={2503.13517},
      archivePrefix={arXiv},
      primaryClass={cs.CL},
      url={https://arxiv.org/abs/2503.13517}, 
}

@misc{lu2024aiscientistfullyautomated,
      title={The AI Scientist: Towards Fully Automated Open-Ended Scientific Discovery}, 
      author={Chris Lu and Cong Lu and Robert Tjarko Lange and Jakob Foerster and Jeff Clune and David Ha},
      year={2024},
      eprint={2408.06292},
      archivePrefix={arXiv},
      primaryClass={cs.AI},
      url={https://arxiv.org/abs/2408.06292}, 
}

@article{Bommasani2021FoundationModels,
title={On the Opportunities and Risks of Foundation Models},
author={Rishi Bommasani and Drew A. Hudson and Ehsan Adeli and Russ Altman and Simran Arora and Sydney von Arx and Michael S. Bernstein and Jeannette Bohg and Antoine Bosselut and Emma Brunskill and Erik Brynjolfsson and S. Buch and Dallas Card and Rodrigo Castellon and Niladri S. Chatterji and Annie S. Chen and Kathleen A. Creel and Jared Davis and Dora Demszky and Chris Donahue and Moussa Doumbouya and Esin Durmus and Stefano Ermon and John Etchemendy and Kawin Ethayarajh and Li Fei-Fei and Chelsea Finn and Trevor Gale and Lauren E. Gillespie and Karan Goel and Noah D. Goodman and Shelby Grossman and Neel Guha and Tatsunori Hashimoto and Peter Henderson and John Hewitt and Daniel E. Ho and Jenny Hong and Kyle Hsu and Jing Huang and Thomas F. Icard and Saahil Jain and Dan Jurafsky and Pratyusha Kalluri and Siddharth Karamcheti and Geoff Keeling and Fereshte Khani and O. Khattab and Pang Wei Koh and Mark S. Krass and Ranjay Krishna and Rohith Kuditipudi and Ananya Kumar and Faisal Ladhak and Mina Lee and Tony Lee and Jure Leskovec and Isabelle Levent and Xiang Lisa Li and Xuechen Li and Tengyu Ma and Ali Malik and Christopher D. Manning and Suvir P. Mirchandani and Eric Mitchell and Zanele Munyikwa and Suraj Nair and Avanika Narayan and Deepak Narayanan and Benjamin Newman and Allen Nie and Juan Carlos Niebles and Hamed Nilforoshan and J. F. Nyarko and Giray Ogut and Laurel Orr and Isabel Papadimitriou and Joon Sung Park and Chris Piech and Eva Portelance and Christopher Potts and Aditi Raghunathan and Robert Reich and Hongyu Ren and Frieda Rong and Yusuf H. Roohani and Camilo Ruiz and Jack Ryan and Christopher R'e and Dorsa Sadigh and Shiori Sagawa and Keshav Santhanam and Andy Shih and Krishna Parasuram Srinivasan and Alex Tamkin and Rohan Taori and Armin W. Thomas and Florian Tram{\`e}r and Rose E. Wang and William Wang and Bohan Wu and Jiajun Wu and Yuhuai Wu and Sang Michael Xie and Michihiro Yasunaga and Jiaxuan You and Matei A. Zaharia and Michael Zhang and Tianyi Zhang and Xikun Zhang and Yuhui Zhang and Lucia Zheng and Kaitlyn Zhou and Percy Liang},
journal={ArXiv},
year={2021},
url={https://crfm.stanford.edu/assets/report.pdf}
}

@inproceedings{
   zhang2025aflow,
   title={{AF}low: Automating Agentic Workflow Generation},
   author={Jiayi Zhang and Jinyu Xiang and Zhaoyang Yu and Fengwei Teng and Xiong-Hui Chen and Jiaqi Chen and Mingchen Zhuge and Xin Cheng and Sirui Hong and Jinlin Wang and Bingnan Zheng and Bang Liu and Yuyu Luo and Chenglin Wu},
   booktitle={The Thirteenth International Conference on Learning Representations},
   year={2025},
   url={https://openreview.net/forum?id=z5uVAKwmjf}
}

@article{he2025llm,
  title={LLM-Based Multi-Agent Systems for Software Engineering: Literature Review, Vision, and the Road Ahead},
  author={He, Junda and Treude, Christoph and Lo, David},
  journal={ACM Transactions on Software Engineering and Methodology},
  volume={34},
  number={5},
  pages={1--30},
  year={2025},
  publisher={ACM New York, NY}
}

@article{murugesan2025rise,
  title={The rise of agentic AI: implications, concerns, and the path forward},
  author={Murugesan, San},
  journal={IEEE Intelligent Systems},
  volume={40},
  number={2},
  pages={8--14},
  year={2025},
  publisher={IEEE}
}

@misc{meta-llm-serving,
title={Scaling Large Language Model Serving Infrastructure at Meta},
author={Ye Qi},
booktitle={QCon San Francisco},
year={2025},
note={https://www.infoq.com/presentations/llm-meta/}
}

@inproceedings{stojkovic2025tapas,
  title={Tapas: Thermal-and power-aware scheduling for LLM inference in cloud platforms},
  author={Stojkovic, Jovan and Zhang, Chaojie and Goiri, {\'I}{\~n}igo and Choukse, Esha and Qiu, Haoran and Fonseca, Rodrigo and Torrellas, Josep and Bianchini, Ricardo},
  booktitle={Proceedings of the 30th ACM International Conference on Architectural Support for Programming Languages and Operating Systems, Volume 2},
  pages={1266--1281},
  year={2025}
}

@article{miao2023towards,
  title={Towards efficient generative large language model serving: A survey from algorithms to systems},
  author={Miao, Xupeng and Oliaro, Gabriele and Zhang, Zhihao and Cheng, Xinhao and Jin, Hongyi and Chen, Tianqi and Jia, Zhihao},
  journal={ACM Computing Surveys},
  year={2023},
  publisher={ACM New York, NY}
}

@inproceedings{wu2024loongserve,
  title={Loongserve: Efficiently serving long-context large language models with elastic sequence parallelism},
  author={Wu, Bingyang and Liu, Shengyu and Zhong, Yinmin and Sun, Peng and Liu, Xuanzhe and Jin, Xin},
  booktitle={Proceedings of the ACM SIGOPS 30th Symposium on Operating Systems Principles},
  pages={640--654},
  year={2024}
}

@inproceedings{gao2024cost,
  title={$\{$Cost-Efficient$\}$ large language model serving for multi-turn conversations with $\{$CachedAttention$\}$},
  author={Gao, Bin and He, Zhuomin and Sharma, Puru and Kang, Qingxuan and Jevdjic, Djordje and Deng, Junbo and Yang, Xingkun and Yu, Zhou and Zuo, Pengfei},
  booktitle={2024 USENIX Annual Technical Conference (USENIX ATC 24)},
  pages={111--126},
  year={2024}
}

@article{wang2024burstgpt,
  title={Burstgpt: A real-world workload dataset to optimize llm serving systems},
  author={Wang, Yuxin and Chen, Yuhan and Li, Zeyu and Kang, Xueze and Tang, Zhenheng and He, Xin and Guo, Rui and Wang, Xin and Wang, Qiang and Zhou, Amelie Chi and others},
  journal={arXiv preprint arXiv:2401.17644},
  year={2024}
}

@inproceedings{jain2025performance,
  title={Performance Aware LLM Load Balancer for Mixed Workloads},
  author={Jain, Kunal and Parayil, Anjaly and Mallick, Ankur and Choukse, Esha and Qin, Xiaoting and Zhang, Jue and Goiri, {\'I}{\~n}igo and Wang, Rujia and Bansal, Chetan and R{\"u}hle, Victor and others},
  booktitle={Proceedings of the 5th Workshop on Machine Learning and Systems},
  pages={19--30},
  year={2025}
}

@article{zhao2025llm,
  title={Llm app store analysis: A vision and roadmap},
  author={Zhao, Yanjie and Hou, Xinyi and Wang, Shenao and Wang, Haoyu},
  journal={ACM Transactions on Software Engineering and Methodology},
  volume={34},
  number={5},
  pages={1--25},
  year={2025},
  publisher={ACM New York, NY}
}

@inproceedings{wang2021faasnet,
  title={$\{$FaaSNet$\}$: Scalable and fast provisioning of custom serverless container runtimes at alibaba cloud function compute},
  author={Wang, Ao and Chang, Shuai and Tian, Huangshi and Wang, Hongqi and Yang, Haoran and Li, Huiba and Du, Rui and Cheng, Yue},
  booktitle={2021 USENIX Annual Technical Conference (USENIX ATC 21)},
  pages={443--457},
  year={2021}
}

@article{agrawal2024vidur,
  title={Vidur: A large-scale simulation framework for llm inference},
  author={Agrawal, Amey and Kedia, Nitin and Mohan, Jayashree and Panwar, Ashish and Kwatra, Nipun and Gulavani, Bhargav S and Ramjee, Ramachandran and Tumanov, Alexey},
  journal={Proceedings of Machine Learning and Systems},
  volume={6},
  pages={351--366},
  year={2024}
}

@inproceedings{wangopenchat,
  title={OpenChat: Advancing Open-source Language Models with Mixed-Quality Data},
  author={Wang, Guan and Cheng, Sijie and Zhan, Xianyuan and Li, Xiangang and Song, Sen and Liu, Yang},
  booktitle={The Twelfth International Conference on Learning Representations}
}

@inproceedings{cohan2018discourse,
  title={A Discourse-Aware Attention Model for Abstractive Summarization of Long Documents},
  author={Cohan, Arman and Dernoncourt, Franck and Kim, Doo Soon and Bui, Trung and Kim, Seokhwan and Chang, Walter and Goharian, Nazli},
  booktitle={Proceedings of the 2018 Conference of the North American Chapter of the Association for Computational Linguistics: Human Language Technologies, Volume 2 (Short Papers)},
  pages={615--621},
  year={2018}
}

@misc{hpc_wire,
    title={"AWS Delivers the AI Heat: Project Rainier and GenAI Innovations Lead the Way"},
    author={"HPC Wire"},
    year = {2024},
    month = {Dec},
    note = {https://www.hpcwire.com/2024/12/05/aws-delivers-the-ai-heat-project-rainier-and-genai-innovations-lead-the-way/}
}

@misc{top500,
    title={"Top 500 Supercomputing List"},
    author={"Top500"},
    year={"2025"},
    month={"June"},
    note={"https://www.top500.org/system/180236/"}
}

@misc{gcp_update,
    title={"From LLMs to image generation: Accelerate inference workloads with AI Hypercomputer"},
    author={"Google Cloud"},
    year={"2025"},
    month={"May"},
    note={"https://cloud.google.com/blog/products/compute/ai-hypercomputer-inference-updates-for-google-cloud-tpu-and-gpu"}
}

@misc{aws_fast_loader,
    title={"Introducing Fast Model Loader in SageMaker Inference: Accelerate autoscaling for your Large Language Models (LLMs)"},
    author={"AWS"},
    year={"2024"},
    month={"Dec"},
    note={"https://aws.amazon.com/blogs/machine-learning/introducing-fast-model-loader-in-sagemaker-inference-accelerate-autoscaling-for-your-large-language-models-llms-part-1/"}
}
